\documentclass[useAMS,usenatbib,usegraphicx]
{mn2e}

\newcommand{\beal}{\begin{align}}
\newcommand{\id}{{\,\rm d}}
\newcommand{\Abst}[1]{\,#1}
\newcommand{\pot}[2]{#1 \times 10^{#2}}
\newcommand{\lesssim}{\mathrel{\hbox{\rlap{\hbox{\lower4pt\hbox{$\sim$}}}\hbox{$<$}}}}
\newcommand{\gtrsim}{\mathrel{\hbox{\rlap{\hbox{\lower4pt\hbox{$\sim$}}}\hbox{$>$}}}} 
\newcommand{\nmax}{n_{\rm max}} 
\newcommand{\nsplit}{n_{\rm split}}

\newcommand{\kB}{k}
\newcommand{\me}{m_{\rm e}}
\newcommand{\Ne}{N_{\rm e}}
\newcommand{\Tg}{T_{\gamma}}
\newcommand{\Te}{T_{\rm e}}
\newcommand{\sigT}{\sigma_{\rm T}}

\newcommand{\pd}{\partial}
\newcommand{\pAb}[2]{\frac{\displaystyle\pd #1}{\displaystyle\pd #2}}

\newcommand{\ion}[2]{{{\rm #1}\,{\sc #2}}}

\newcommand{\change}[1]{{#1}}

\newcommand{\hdist}{-1.0mm}
\newcommand{\vdist}{1.5mm}
\newcommand{\pscale}{0.69}

\usepackage{amsmath}

\title[Cosmological hydrogen recombination] 
{Cosmological hydrogen recombination: populations of the high level sub-states}
\author[]{J. Chluba$^1$\thanks{E-mail: 
jchluba@mpa-garching.mpg.de}, J.~A. Rubi\~no-Mart\'{\i}n$^{2}$\thanks{E-mail: 
jose.alberto.rubino@iac.es} and R.A. Sunyaev$^{1,3}$\\
$^{1}$ Max-Planck Institut f\"ur Astrophysik, Karl-Schwarzschild-Str. 1,
D-85740 Garching, Germany\\
$^{2}$ Instituto de Astrof\'isica de Canarias, C/V\'ia L\'actea s/n, 
    E-38200 Tenerife, Spain\\
$^{3}$ Space Research Institute (IKI), Russian Academy of Sciences, Profsoyuznaya 84/32
Moscow, Russia }

\begin{document}

\date{Received **insert**; Accepted **insert**}

\pagerange{\pageref{firstpage}--\pageref{lastpage}}
\pubyear{}

\maketitle

\label{firstpage}

\begin{abstract}
We present results for the spectral distortions of the Cosmic Microwave
Background (CMB) arising due to bound-bound transitions during the epoch of
cosmological hydrogen recombination at frequencies down to $\nu\sim
100\,$MHz. We extend our previous treatment of the recombination problem
\citep{Jose2006} now including the main collisional processes and following
the evolution of all the hydrogen \change{angular momentum} sub-states for up
to 100 shells.
We show that, due to the low baryon density of the Universe, even within the
highest considered shell full statistical equilibrium (SE) is not reached and
that at low frequencies the recombination spectrum is significantly different
when assuming full SE for $n>2$.
We also directly compare our results for the ionization history to the output
of the {\sc Recfast} code, showing that especially at low redshifts rather big
differences arise. In the vicinity of the Thomson visibility function the
electron fraction differs by roughly $-0.6\%$ \change{which affects the
temperature and polarization power spectra by $\lesssim 1\%$}.
Furthermore we shortly discuss the influence of free-free absorption and line
broadening due to electron scattering \change{on the bound-bound recombination
spectrum} and the generation of CMB angular fluctuations due to scattering of
photons within the high shells.

\end{abstract}


\section{Introduction}

The study of the Cosmic Microwave Background (CMB) provides one of the most
powerful tools to test cosmological models and to determine the parameters
describing the Universe.
Nowadays, most of the CMB experiments are focused on the study of the
temperature and polarization angular fluctuations, which have been measured
recently with very high precision by the {\sc Wmap} satellite \citep{Page2006,
Hinshaw2006}, permitting to achieve accuracies in the determination of some of
the key cosmological parameters on the level of a few percent.

The CMB photons which are detected today were essentially last scattered
towards us during the epoch of cosmological recombination, where the
temperature of the Universe had become sufficiently low to permit the
formation of neutral atoms \citep{Zeldovich68,Peebles68}.
During the process of cosmological recombination of the hydrogen and helium
atoms there was a net generation of photons, which introduces distortions to
the CMB blackbody spectrum. Except for the far Wien tail of the CMB, these
distortions are expected to be very small:
the distortions due to bound-bound transitions reach the level of $\Delta
I/I\sim 10^{-7}$ at frequency $\nu\sim 1\,$GHz and are smaller than $\Delta
I/I\sim 10^{-8}$ in the range $10\,\text{GHz}\leq\nu \leq 500\,$GHz
\citep[e.g. see][RMCS06 hereafter]{Jose2006}.
This is mainly due to the extremely large entropy of the Universe (there are
$\sim \pot{1.6}{9}$ photons per baryon), since in the presence of undistorted
blackbody radiation there cannot be much more than one photon per transition
\citep{Liubarskii83}.
The free-bound emission increases the spectral distortions by additional $\sim
30-90$\% \change{and a total $\sim5$ photons per hydrogen atom are released
due to hydrogen recombination in the undistorted ambient blackbody photon
field of the CMB} \citep{Chluba2006b}.
However, a measurement of these tiny deviations from a pure blackbody would
provide another independent way to determine some of the key cosmological
parameters, such as the baryon density or the total matter content. In
addition they in principle are a {\it byproduct} of the accurate treatment of
the recombination history and in particular the time dependence of the free
electron fraction in the Universe, which today is so important for the
theoretical prediction of the CMB temperature and polarization power spectra.

%
The numerical solution of the hydrogen recombination history and the
associated spectral distortions of the CMB requires the integration of a stiff
system of coupled ordinary differential equations, describing the evolution of
the populations of the different hydrogen levels, with extremely high
accuracy, simply because the amplitude of the distortions is so small.
Several authors have computed the distortions due to bound-bound transitions
for a multi-level hydrogen atom with different assumptions and simplifications
\citep[][RMCS06]{Dubrovich1975,Beigman1978,Rybicki93,DubroVlad95,Dubrovich2004,Kholu2005,Wong2006}.
Among these the most important simplification is to assume {\it full
  statistical equilibrium}\footnote{i.e. the population of a given level
  $(n,l)$ is determined by $N_{nl} = (2l+1) N_n / n^2$, where $N_n$ is the
  total population of the shell with principle quantum number $n$.} (SE)
  within a given shell for $n>2$ (for a more detailed comparison of the
  different approached see RMCS06 and references therein). With this
  assumption the recombination problem becomes significantly simpler, since
  for each additional shell only one more equation has to be treated, and
  \citet{Kholu2005} already presented results for the hydrogen recombination
  spectrum due bound-bound transitions in the frequency range
  $1\,\text{GHz}\leq\nu\leq 1000\,\text{GHz}$ including up to 160 shells
  \change{within} this approximation.

In our previous work (RMCS06) the spectral distortions of the CMB spectrum
arising due to bound-bound transitions and the 2s two-photon decay during the
epoch of cosmological hydrogen recombination were obtained in the frequency
range $1\,\text{GHz}\leq\nu\leq 3500\,\text{GHz}$.
There the complete set of ordinary differential equations was solved for the
cosmological hydrogen recombination problem including up to 30 shells, with
{\it all} the energetically degenerate angular momentum sub-levels treated
separately.
It was shown that assuming SE for shells $n>2$ leads to significant
differences in the cosmological hydrogen recombination spectrum for the
Balmer, Paschen and Brackett series.
Furthermore, it was argued that collisions probably can only become important
for shells well above $n\sim 25-30$, so that assuming SE for the lower shells
is not justified, but a detailed examination of this problem has not been
performed until now.
%

In this paper, we present detailed results for the hydrogen recombination
spectrum due to bound-bound transition at low frequencies down to $\nu\sim
100\,$MHz (see Fig.~\ref{fig:DI_results}). To this end, we have solved the
complete set of equations for a multi-level hydrogen atom with 100 shells,
i.e. we are following the populations of 5050 separate hydrogen states.
Furthermore, we include the main collisional processes into our calculations
(for more details see Sect.~\ref{sec:coll_details}) and discuss deviations
from full SE within the high shells. We also examine the influence of the
free-free process (Sect.~\ref{sec:freefree}) at low frequencies, the line
broadening due to electron scattering (Sect.~\ref{sec:el_sc}) and the
generation of CMB angular fluctuations by scattering of photons within the
high shells (Sect.~\ref{sec:more}).
For the treatment of the multi-level hydrogen atom we basically follow the
same procedure as described in RMCS06, with some modification which are
explained in Sect.~2.

Several authors have worked on percent level corrections to the ionization
history arising from different previously neglected physical processes
\citep{Dubrovich2005, Chluba2006, Kholu2006, Novosyadlyj2006}. It was pointed
out that these corrections lead to differences in the theoretical predictions
for the CMB temperature and polarization power spectra on a similar level and
therefore should be taken into account when analyzing future high precision
data.
Recently, \citet{Lewis2006} have made some first steps towards quantifying the
possible impact of percent level correction to the ionization history on the
estimation of cosmological parameters, again showing that in the era of
precision cosmology more efforts should be made to improve the calculation of
the recombination history.

In RMCS06 it was also pointed out that there are corrections to the ionization
history on the level of percent arising due to the detailed treatment of the
angular momentum sub-states of the hydrogen atom.
With computations including up to 100 shells, one reaches the point to
directly compare the obtained ionization history with the output of the {\sc
  Recfast} code, which was developed to represents the computations based on a
300-level hydrogen atom \citep{SeagerRecfast1999}\change{, but under the
  assumption of SE for $n>2$.}
We shall discuss our results for the obtained electron recombination history
as compared to the {\sc Recfast} output in Sect.~\ref{sec:el_hist} and show
that especially at low redshifts rather big differences arise.
However, still more work has to be done to reach percent level accuracy
\change{for the ionization history}.

\change{For all the computation presented in this work the following values of the
cosmological parameters were adopted: $\Omega_{\rm b}=0.0444$, $\Omega_{\rm
  tot}=1$, $\Omega_{\rm m}=0.2678$, $\Omega_{\Lambda}=0.7322$, $Y_{\rm
  p}=0.24$, $h=0.71$ and $T_0=2.725\,$K.
}

%
\section{Computational issues}
In general we use the same approach and definitions as in RMCS06.  However, in
order to solve the full problem for $\nmax\geq 30$ we had to improve the
computational treatment significantly. Therefore we modified one of our
previous codes to include the most important collisional processes and made
some more simplifying assumptions, which we shall briefly discuss in this
Section.

\subsection{Inclusion of collisions}
\label{sec:coll_details}
\begin{figure}
\centering 
\includegraphics[width=\columnwidth]{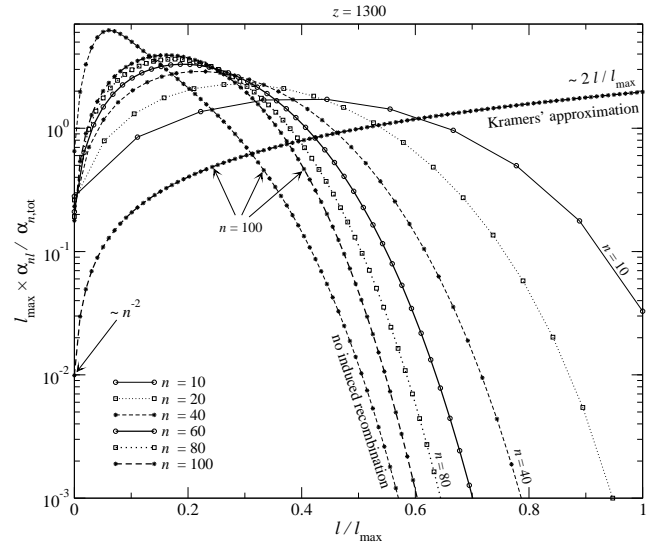}
\caption{$l$-dependence of the recombination coefficient, $\alpha_{nl}$, at
$z=1300$ for different shells. The curves have been re-scaled by the {\it
total} recombination coefficient, $\alpha_{n,\rm tot}=\sum_l \alpha_{nl}$, and
multiplied by $l_{\rm max}=n-1$ such that the 'integral' over $\xi=l/l_{\rm
max}$ becomes unity. Also the results obtained within the Kramers'
approximation, i.e. $\alpha_{nl}^{\rm K}={\rm const}\times[2l+1]$, and without
the inclusion of stimulated recombination for $n=100$ are presented.}
\label{fig:Rci_Rci_tot}
\end{figure}

It was pointed out in RMCS06 that the assumption of SE is equivalent to {\it
instantaneous redistribution} of the electrons within a given
shell. Especially for hydrogen levels $(n, l)$ with large angular momentum
quantum number ($0\ll l\leq n-1$) this assumption changes the populations
strongly, because for high shells due to the $l$-dependence of the
\change{recombination} Gaunt-factor direct recombinations to these levels
become very unlikely.
\change{This point is illustrated in Fig.~\ref{fig:Rci_Rci_tot}, where we show
  the $l$-dependence of the recombination coefficients $\alpha_{nl}$ for
  different shells\footnote{Here $\alpha_{nl}$ is defined such that $N_{\rm
  e}\,N_{\rm p}\, \alpha_{nl}$ is the change of the population of level
  $(n,l)$ due to direct recombinations per second, where $N_{\rm e}$ and
  $N_{\rm p}$ are the free electron and proton number densities,
  respectively.}. One can clearly see, that for large $n$ most of the
  electrons recombine to states with $l/l_{\rm max}\sim 0.1-0.2$.
}
Therefore in the full problem electrons mainly reach \change{levels with large
  $n$ and $l$} connecting to other levels via radiative and collisional
transitions, where the former are very inefficient, since they proceeds via
many intermediate states, and latter, due to the large entropy of the
Universe, are rather slow.
\change{However, due to the $\Delta l=\pm 1$ restriction for radiative dipole
transitions, levels with large $n$ and $l$ quantum numbers will depopulate
rather slowly via low frequency transitions which leads to a delicate
interplay between all the rates connecting to the neighboring levels.}

\change{In Fig.~\ref{fig:Rci_Rci_tot} for $n=100$ we also give the
recombination coefficient within the Kramers' approximation. Here the
dependence on the recombination Gaunt-factor is neglected such that the
$l$-dependence of the recombination coefficient only enters due to the fact
that {\it one single} electron can find $(2l+1)$ sub-levels to be captured
to. Hence, within the Kramers' approximation electrons preferentially
recombine to states with large $l$.
On the other hand, it is clear that within the Kramers' approximation the
photoionization process is also more effective for high $l$-states, which
implies that after any disturbance for these levels full equilibrium with the
continuum is reached on a much shorter time-scale. Since here we are looking
for {\it tiny} deviations from full equilibrium therefore it is important to
include the Gaunt-factors into the computations and to account for collisions
which help changing the populations of high $l$-states.

From Fig.~\ref{fig:Rci_Rci_tot} one can also see that explicitly neglecting
{\it induced recombinations}, but including the Gaunt-factor, significantly
alters the $l$-dependence of recombination coefficient for large $n$. In
addition the total recombination rates for high shells strongly decreases for
this case.}
Below, we now outline which collisional processes were accounted for in the
present paper.

\subsubsection{$l$-changing collisions}
One of the most important differences with respect to our previous
computations (RMCS06) is that we now also include collisions of the form
$(n,l)\rightarrow (n, l\pm 1)$.
These are important in order to {\it restore} and {\it maintain} full
statistical equilibrium (SE) within a given shell and from simple estimates
\citep{Pengelly1964} for the conditions like in our Universe one expects that
within the redshift range $500\leq z\leq 2500$ these start contributing for
shells with principal quantum numbers $n\gtrsim 25-30$.
The corresponding coupling term for the population $N_{nl}$ of level $(n, l)$
to the populations $N_{nl'}$ of level $(n, l')$ with $l'=l\pm1$ due to the
collision with a particle of species i is then given by
\begin{equation}
\label{eq:Coll}
\left. \pAb{N_{nl}}{t}\right|_{\rm i}^{l-\rm coll} 
=
\left[ \frac{(2l+1)}{(2l'+1)}\frac{N_{nl'}}{N_{nl}}-1 \right] N_{nl} C_{nl\rightarrow nl'}
\Abst{,}
\end{equation}
We calculate the collision rates $C_{nl\rightarrow nl'}$ according to
\citet{Brocklehurst1971}. 

Since the energy of the bound electron is not changing, in contrast to
$n$-changing collisions the mass of the projectile is more important than its
velocity. Therefore the $l$-changing collision rates due to proton impact are
usually larger than for electron impact, but we included both into our
calculations. Also collisions with \ion{He}{ii} ions were accounted for, but
due to their low number density and the history of \ion{He}{ii} recombination
they only lead to a $\lesssim 10\%$ contribution to the total collision rate
at redshifts $z\gtrsim 1500$. Since the lines due to hydrogen recombination
typically appear at redshifts $z\lesssim 1500$, \ion{He}{ii} collisions
basically do not affect the final results for the bound-bound spectral
distortion.

In order to speed up our computation we tabulate the $l$-changing collision
rates before starting to solve the system of coupled ordinary differential
equations, assuming that a sufficient approximation to the electron
temperature $\Te$ and electron, proton and \ion{He}{ii} number densities is given by
the results from the {\sc Recfast} code \citep{SeagerRecfast1999}. Since in
general the approximations for the collisional rates are only accurate on the
level of $\sim 10\%-50\%$ this should be a reasonable approach.

One should mention that in principle the $\Delta l=\pm 1$ restriction does not
apply for $l$-changing collisions and coupling terms with $|\Delta l| > 1$
could be important, but smaller due to the decrease of the cross section with
growing $\Delta l$.  These should help to bring the populations for a given
shell faster into full SE than in the calculations presented here.  However,
we neglected those terms for the moment.

\subsubsection{$n$-changing collisions}
Energy or $n$-changing collisions of the type $(n,l)\rightarrow (n', l\pm 1)$
with $n\neq n'$ are usually less important than $l$-changing collisions
\citep[e.g. see comments in Sect. 3.2.4 of][]{Hummer1987}. But since here we
are discussing highly excited electron states in the hydrogen atom we decided
to include collisional excitation and de-excitation as well and check their
influence on the recombination spectrum. However, we are not including
collisional ionization or three-body recombination, since the relevance of
these processes is expected to be even smaller.

As mentioned above, for energy-changing collisions the velocity of the
projectile is important and hence the $n$-changing collision rate is
dominated by the contribution from electron impact. We shall only consider
collisional excitation and de-excitation by electrons, but neglect the
corresponding processes induced by protons, \ion{He}{ii} ions or neutral hydrogen
atoms.

For collisional processes a vast amount of literature can be found, all based
on different assumptions and approximations. Here we decided to use the simple
approximations as given by \citet{Rege1962}. These likely are only accurate to
within a factor of $\sim 2-10$ probably even up to $\sim 100$, but they are
very fast for numerical evaluations and widely used in computations of stellar
atmospheres.
The corresponding coupling term for the population $N_{nl}$ of level $(n, l)$
to the populations $N_{n'l'}$ of level $(n', l')$ due to the collision with an
electron is then given by
\begin{equation}
\label{eq:nColl}
\left. \pAb{N_{nl}}{t}\right|_{\rm e}^{n-\rm coll} 
\!\!=
\left[
\frac{(2l+1)}{(2l'+1)}\frac{N_{n'l'}}{N_{nl}}\,e^{-\frac{\Delta E}{\kB\Te} }-1 
\right]\! 
N_{nl} C_{nl\rightarrow n'l'}
\Abst{,}
\end{equation}
where $\Delta E$ is the energy difference of the considered levels. We
calculate the collisional de-excitation rate $C_{nl\rightarrow n'l'}$ for
electron impact according to formula (18) in the work of \citet{Rege1962}.
Only radiatively allowed transitions are taken into account here. Note that
since collisions are controlled by electrons, the electron temperature appears
in the exponential factor of \eqref{eq:nColl}.


\subsection{Additional comments}
\label{sec:addcomm}
The computations of the photoionization and photorecombination rates involve
1-dimensional integrals over the ambient photon field. These have to be
evaluated many times in order to solve the problem. Since we assume that the
ambient photon field is a pure blackbody, the only coupling to the current
solution arises due to the electron temperature.
For up to $50$ shells we have checked that instead of evaluating both the
photoionization and photorecombination rates at every time it is sufficient to
use the photorecombination rate {\it only} and apply the detailed balance
relation. Since this approximation is possible one can again tabulate the
photorecombination rates using the results for the electron temperature from
the {\sc Recfast} code and then interpolate. In this way one speeds up the
computation by a large factor. For $50$ shells we have compared the results
obtained with and without this simplification and found practically no
difference \change{in the recombination spectrum. This is because at $z\sim
1100- 1400$, i.e. where the lines form, the difference in the electron and
photon temperature is very small ($\Delta T/T \sim 10^{-6}-10^{-5}$).}
One should also mention that we evaluated the photoionization cross section
and radiative rates as described in \citet{StoreyHum1991}, since for $n\gtrsim
50$ the expressions given by \citet{Karzas1961} were numerically unstable. 

In addition it is important to use both relative and absolute error control.
The solvers {\sf D02NBF} and {\sf D02NGF} from the {\sc
Nag}-Library\footnote{see http://www.nag.co.uk/} provide this possibility. We
usually requested $\epsilon\sim 10^{-8}$ relative and $\epsilon\sim
10^{-22}-10^{-23}$ or less absolute error. Furthermore at low redshifts we had
to limit the maximal step-size to $h\sim 0.1-0.5$. We also ran computations
with increased accuracies but found no \change{significant} differences.
With these settings one computation of the recombination spectrum for
$\nmax=\nsplit=100$ took about $6$ days on a $\sim 3\,$GHz single processor
machine.

\section{Results and discussion}

\begin{figure}
\centering 
\includegraphics[width=\columnwidth]{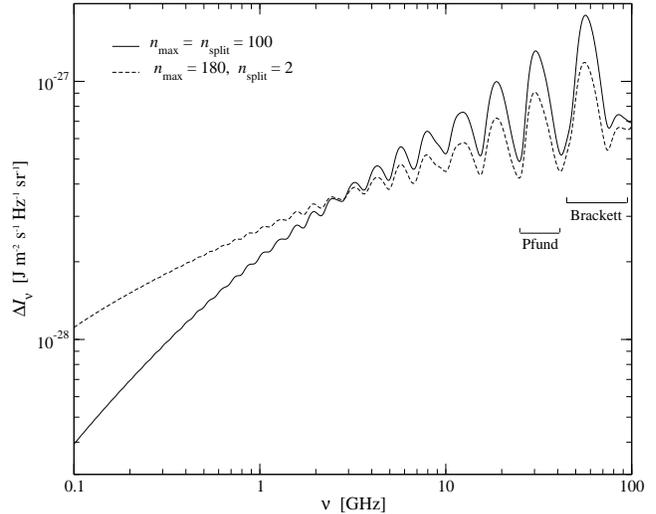}
\caption{The hydrogen bound-bound emission spectrum at low frequencies. In the
full computation ($\nmax=\nsplit=100$) the evolution of all the angular
momentum sub-states was taken into account separately including collisions.
For comparison we also show the results obtained with $\nmax=180$ and
$\nsplit=2$. The contribution due to the 2s two-photon decay is negligible
within the considered range of frequencies and was therefore omitted.
}
\label{fig:DI_results}
\end{figure}
\begin{figure}
\centering 
\includegraphics[width=0.98\columnwidth]{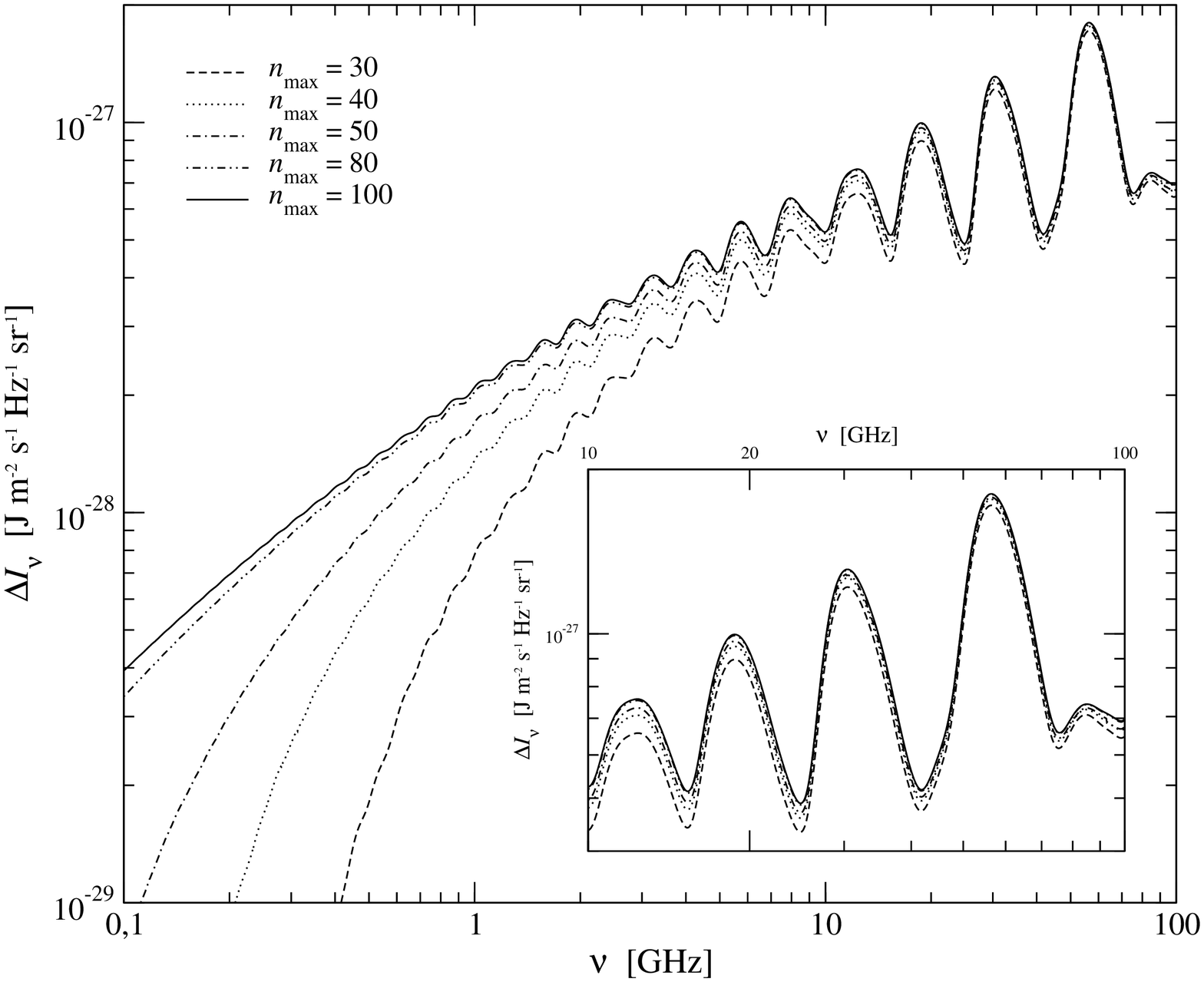}
\caption{The hydrogen bound-bound emission spectrum at low frequencies for
different $\nmax$ and $\nsplit=\nmax$. Also $l$- and $n$-changing collisions
were included.}
\label{fig:DI_conv_full}
\end{figure}
\begin{figure}
\centering 
\includegraphics[width=0.98\columnwidth]{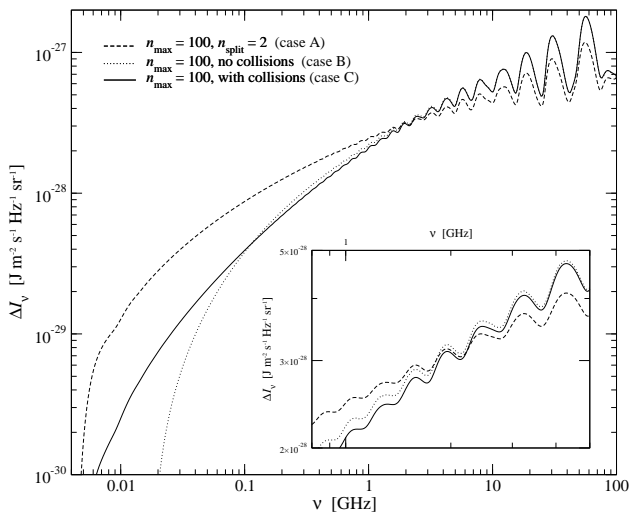}
\caption{Effect of collisions on the hydrogen bound-bound emission spectrum at
low frequencies. Free-free absorption was not taken into account.}
\label{fig:DI_coll_full}
\end{figure}
\subsection{The hydrogen bound-bound emission spectrum at low frequencies}
In Figure~\ref{fig:DI_results} we present the hydrogen recombination spectrum
for bound-bound transitions at low frequencies.
In our full computation ($\nmax=\nsplit=100$) the evolution of {\it all} the
angular momentum sub-states was taken into account separately and $l$- and
$n$-changing collisions were included. In this calculation the evolution of
$5050$ separate hydrogen states in the redshift range of $200\leq z\leq 2100$
was followed.
Due to computational limitations this was the largest calculation we
were able to perform within reasonable time.
For comparison we also show the results obtained with $\nmax=180$ under the
simplifying assumption of SE for shell with $n>2$, i.e. $\nsplit=2$.

For the full calculation ($\nmax=100$ and $\nsplit=100$) within the frequency
range $1\,\text{GHz}\leq\nu\leq 100\,\text{GHz}$ one can clearly see the low
contrast features, which were also discussed in RMCS06.  As expected, here due
to the inclusion of more shells the level of the distortions is higher than in
RMCS06, but at $\nu\lesssim 2\,$GHz they do not reach the level obtained in
the case $\nmax=180$ and $\nsplit=2$.
At $1\,\text{GHz}$ these two solutions differ by a factor of $\sim 1.4$ (see
Section~\ref{sec:collisions} for more discussion).
Below $\sim 1\,$GHz the spectral distortions become practically featureless.
In the frequency range $1\,\text{GHz}\lesssim\nu\lesssim 10\,\text{GHz}$ the
slope of the spectrum is $\beta\sim 0.46$ and therefore slightly steeper than
the one obtained in the calculations of \citet{Kholu2005} and our $\nmax=180$
and $\nsplit=2$ case, i.e. $\beta\sim 0.35$.

\subsubsection{Convergence with $\nmax$}
\label{sec:convergence}
To estimate the convergence of the results for the hydrogen recombination
spectrum at low frequencies we computed the distortions for different $\nmax$
(see Fig.~\ref{fig:DI_conv_full}). At $\nu\sim 1\,$GHz for $\nmax=100$ the
spectral distortion is a factor of $\sim 2.6$ larger than for $\nmax=30$. One
can see that at $\nu\geq 1\,$GHz the distortion is practically not changing
anymore when including more than $\sim 80$ shells.
Within our assumption the results should be converged on a level of $\sim
10-20$\% at frequencies $0.1\,\text{GHz}\leq\nu\leq 1\,$GHz and to better than
1\% for $\nu\gtrsim 1\,$GHz. We also checked the convergence for up to
$\nmax=180$ but with $\nsplit=2$ and reached to the same conclusion.

\subsubsection{The influence of collisions}
\label{sec:collisions}
Here we wish to understand how the inclusion of collisions affects the
hydrogen recombination spectrum. In Figure~\ref{fig:DI_coll_full} we compare
the results for the spectral distortions down to very low frequencies obtained
for $\nmax=100$ in three different calculations: case $\mathcal{A}$:
$\nsplit=2$; case $\mathcal{B}$: $\nsplit=\nmax$, no collisions; case
$\mathcal{C}$: $\nsplit=\nmax$, with collisions.

First one can clearly see the difference in the slope of the distortions in
the range $0.1\,\text{GHz}-10\,$GHz for the cases $\mathcal{B}$ and
$\mathcal{C}$ in comparison with case $\mathcal{A}$. At frequencies below
$\nu\sim 2\,$GHz the solution for case $\mathcal{A}$ lies well above the other
two solutions. The solution for case $\mathcal{B}$ cuts off at $\nu\sim
200\,$MHz, while the ones for the cases $\mathcal{A}$ and $\mathcal{C}$ reach
down to $\nu\sim 45\,$MHz.
This shows that {\it without collisions} transitions with $\Delta n\ll
n$ at large $n$ are disfavored and electrons reaching states $l\ll n$ to a
large fraction depopulate via transitions with $\Delta n\sim n$, which
increases the emission at higher frequencies (see also RMCS06).
In addition one can conclude that for case $\mathcal{A}$ and $\mathcal{C}$ the
cascade of electrons within the high $l$-states starts to become similarly
important.
However, since the populations of the very low shells ($n\leq 20-30$) are not
significantly altered by collision, at low frequencies the distortions for
case~$\mathcal{C}$ cannot reach the same level as in the case~$\mathcal{A}$,
simply because the net radiative rates and population of the high
shells critically depend on the cascade to lower shells.

We have checked, that increasing the rates for $l$-changing collisions by a
large factor ($\sim 10^4-10^6$) we can recover the solution like for
$\nsplit=2$. Furthermore we have performed \change{additional} calculations
with $\nmax>100$ and $\nsplit=100$ in order to check how this affects the
recombination spectrum. We found that it only influences the distortions close
to the lowest frequencies and still there the level of the case $\mathcal{A}$
is not reached.
Therefore we conclude that the difference in the slopes of the recombination
spectrum for case $\mathcal{A}$ and $\mathcal{C}$ at low frequencies will
remain, even for $\nmax=\nsplit>100$.

We also studied how important in particular $n$-changing collisions are by
performing computations only including $l$-changing collisions. In this case
the spectrum practically remained unchanged (differences $\lesssim 1\%$ at the
very low frequencies). In addition we ran computations artificially increasing
the rate of $n$-changing collisions by a factor of 100. Here we found that the
recombination emission at low frequencies was \change{slightly} reduced (at
most by factor of $\sim 2$ for $\nu < 100\,$MHz in this case).
This is due to the fact that in this case the net rate connecting two
\change{levels with large $n$} includes a sizeable non-radiative
contribution. One would expect that this, in addition to free-free absorption,
may lead to a reduction of emission at very low frequencies, simply because
$n$-changing collisions should dominate over radiative transitions for some
large value of $n$, but it seems that this will only happen for $\nmax >100$
and correspondingly at \change{even lower frequencies than considered here}.

\begin{figure}
\centering 
\includegraphics[width=0.98\columnwidth]{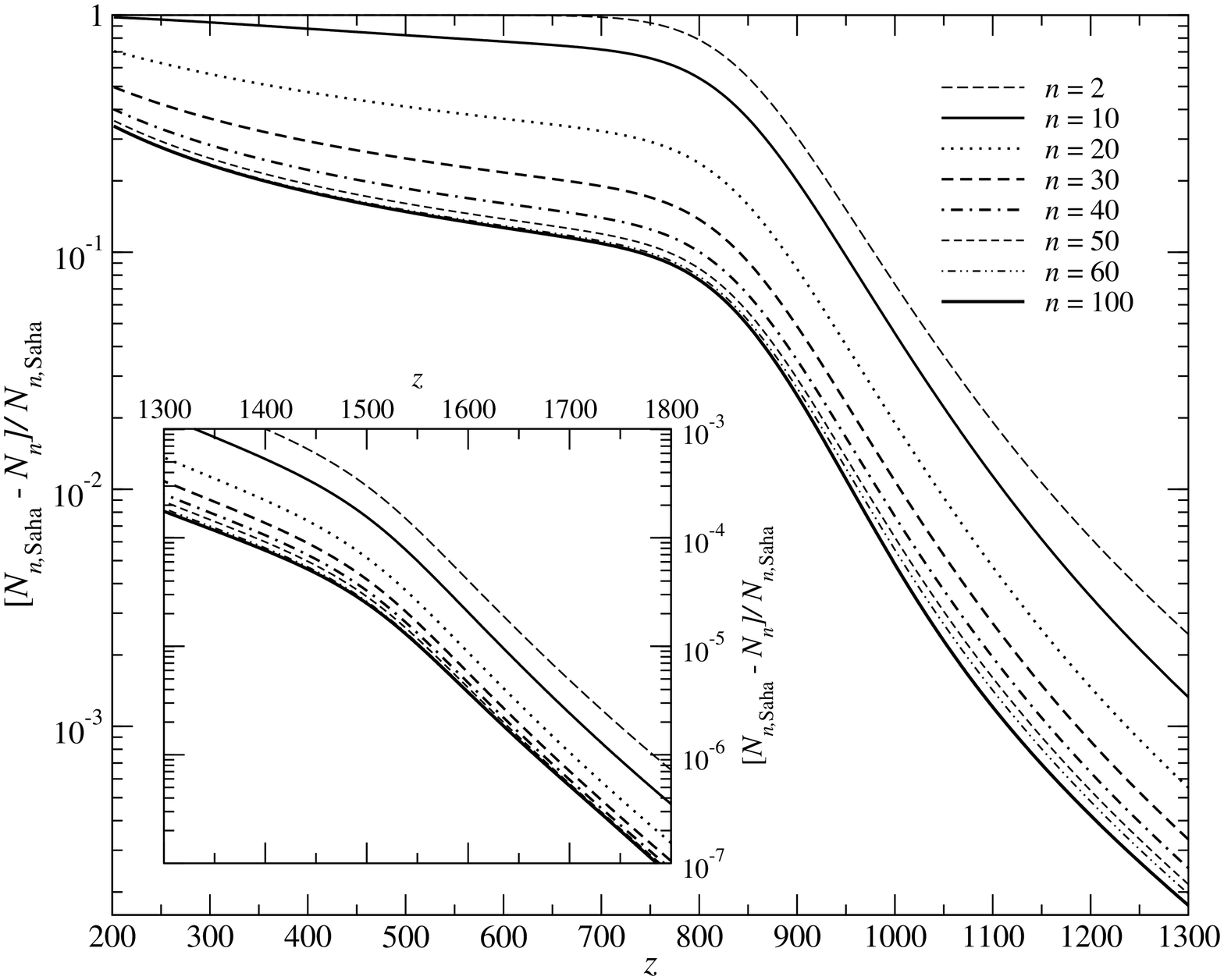}
\\
\includegraphics[width=0.98\columnwidth]{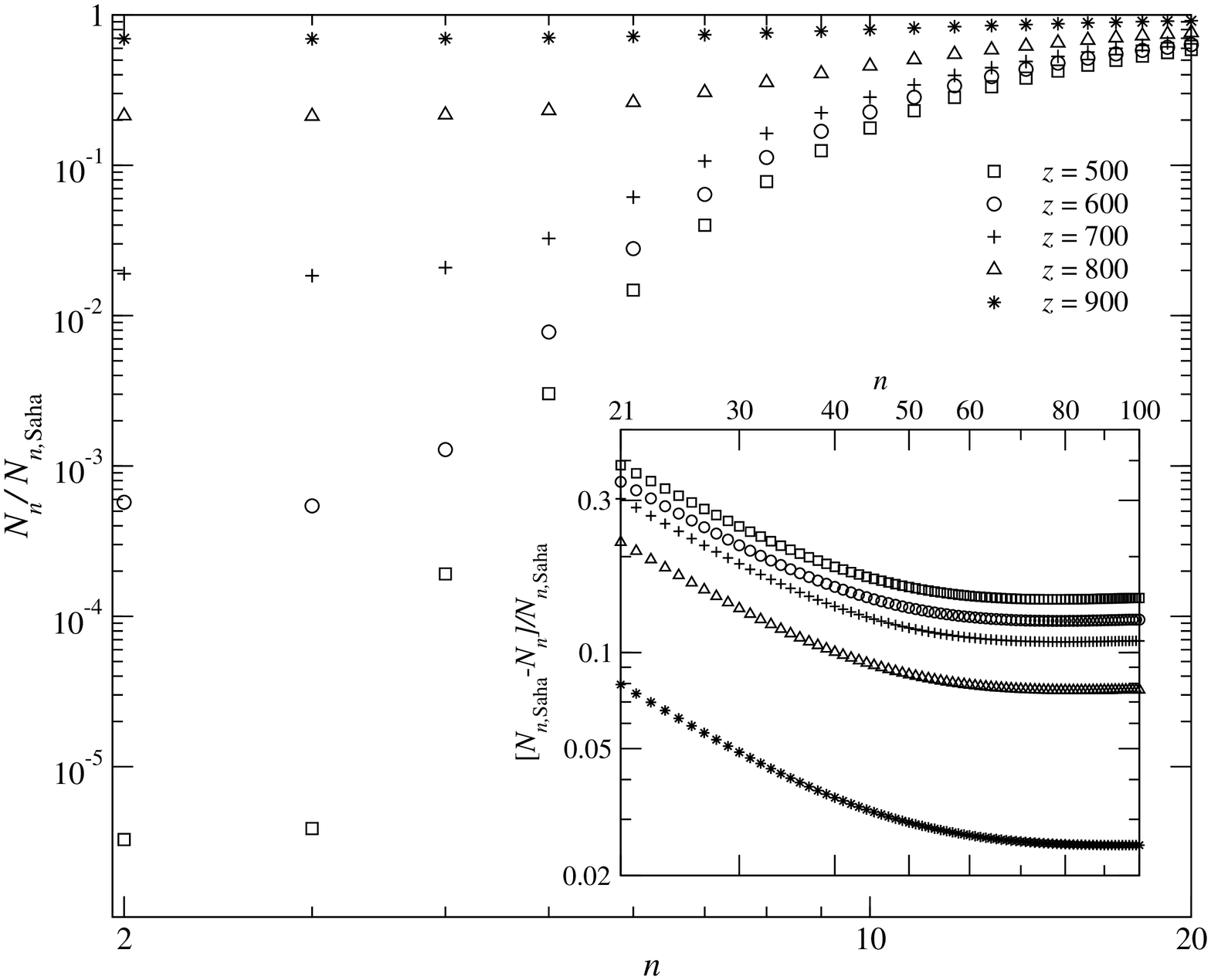}
\\
\includegraphics[width=0.98\columnwidth]{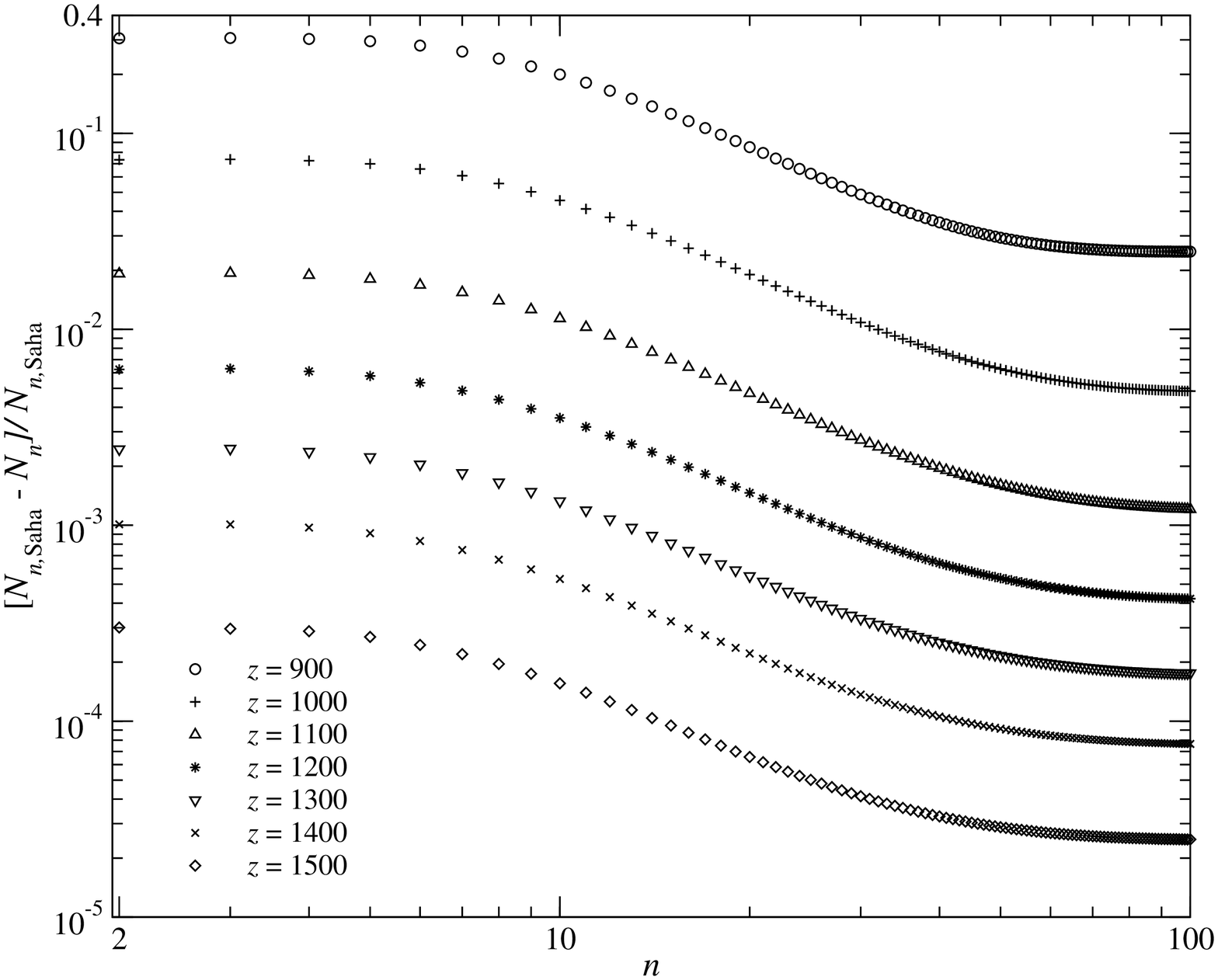}
\caption{Departures of the total populations from Saha: $N_n$ denotes the
total population of shell $n$ as obtain in our 100 shell computation with
$\nmax=\nsplit$ including collisions. $N_{n, \rm Saha}$ is the expected
Saha-population relative to the continuum.}
\label{fig:DN_Saha}
\end{figure}
\subsection{Populations of the levels}
\label{sec:pops}
In this Section we discuss the structure of the hydrogen level populations in
some more detail. At redshifts $z\leq 1500$ the ground state population always
is extremely far from Saha-equilibrium with the free electrons. We
therefore omitted it in the following.

Figure~\ref{fig:DN_Saha} shows the departures of the total populations for
given $n$ from the expected Saha population relative to the continuum.  At
high redshifts ($z\gtrsim 1800$), for all shells with $n\geq 2$ the departures
from Saha are $\lesssim 10^{-7}-10^{-6}$.
At $z\sim 1500$ the departures reach a level of $\sim 10^{-5}-10^{-4}$ and for
$z\lesssim 1500$ all the curves become flatter, because from there on
recombination is controlled by the escape of photons from the Lyman-$\alpha$
transition.
At $z\sim 800$ the deviations from Saha exceed $\sim 10\%$ for all the
considered shells and below that redshift they only change within a factor of
$\sim 2-5$ until redshift $z=200$.
In general one can also see that for shells with $n\lesssim 10$ the departures
from Saha are typically larger than for higher $n$ and that below $z\sim 800$
the total populations of these levels are overestimated by many orders of
magnitude when applying the Saha-relation with respect to the continuum.
Note that during the whole epoch of recombination the total population of any
shell is {\it smaller} than the expected Saha value.

\begin{figure*}
\centering 
\includegraphics[width=\pscale\columnwidth]{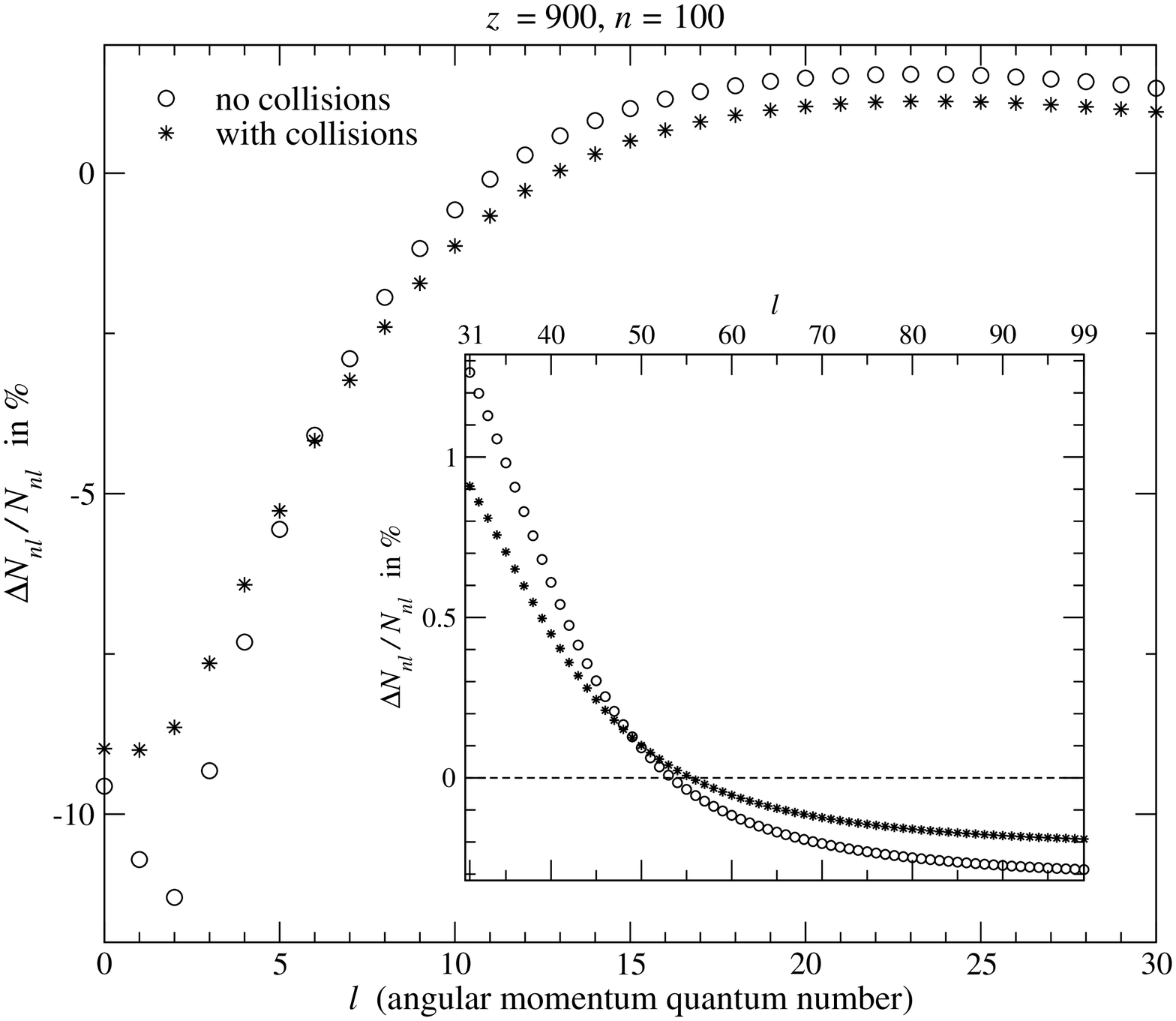}
\hspace{\hdist}
\includegraphics[width=\pscale\columnwidth]{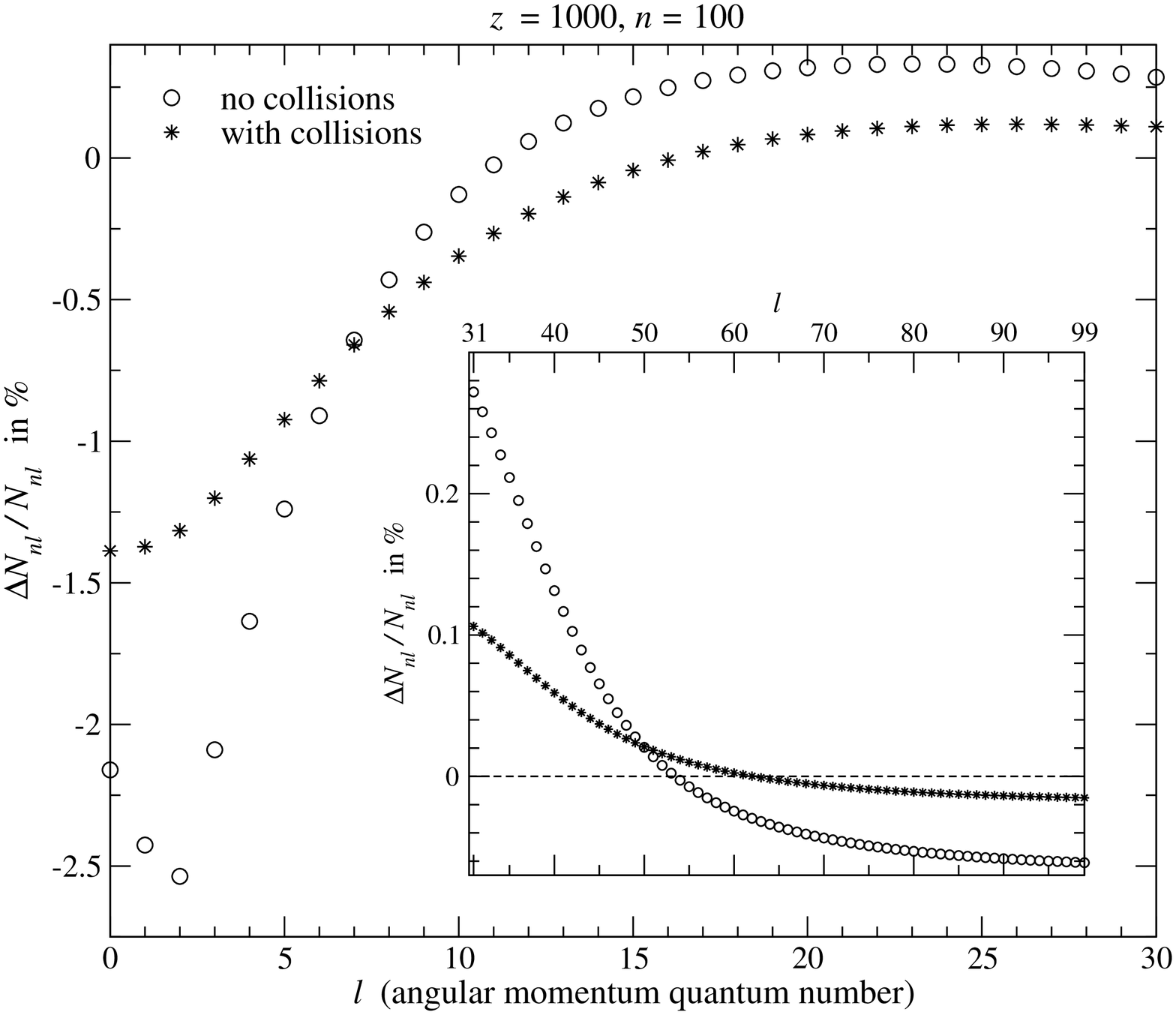}
\hspace{\hdist}
\includegraphics[width=\pscale\columnwidth]{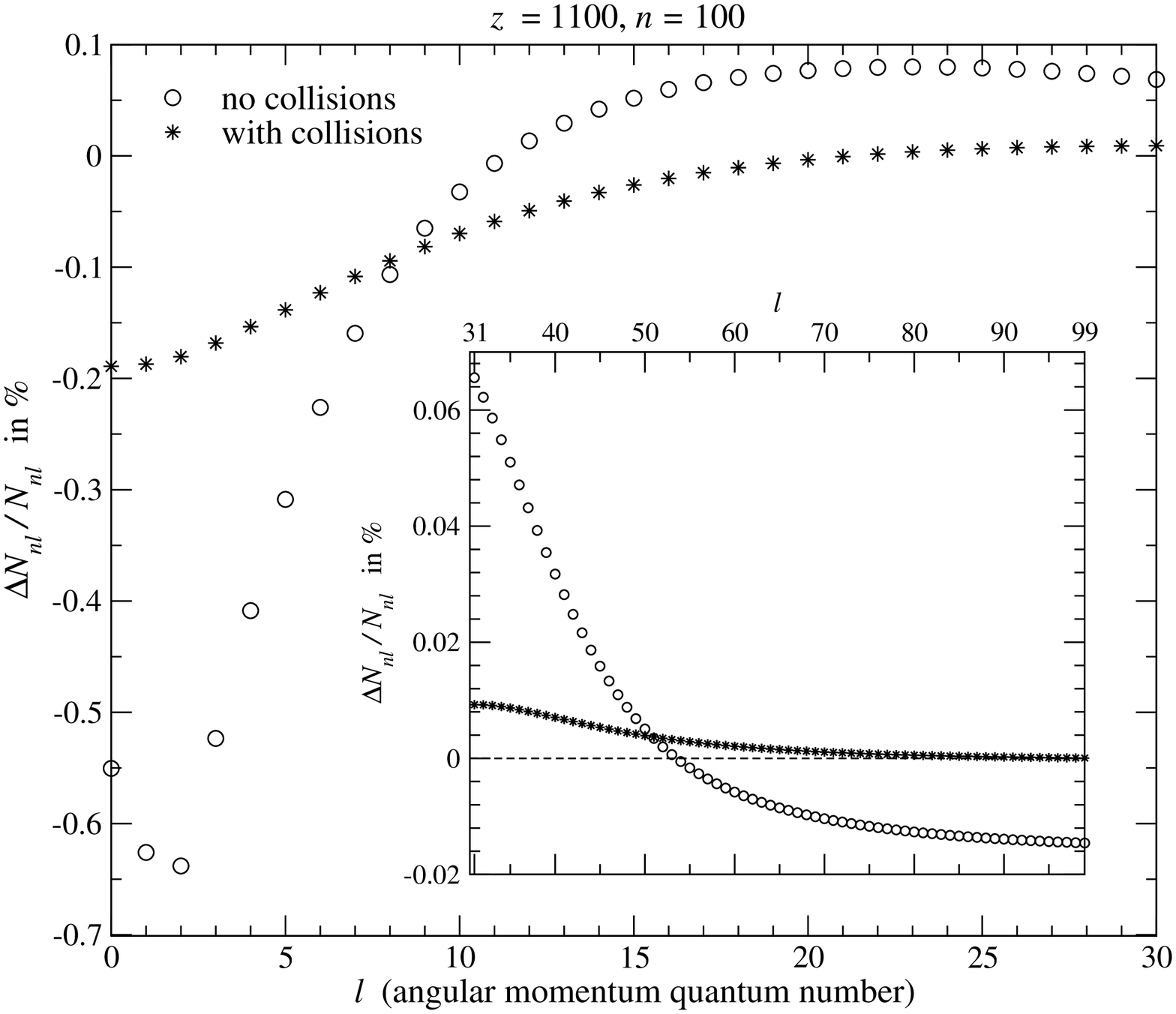}
\\[\vdist]
\includegraphics[width=\pscale\columnwidth]{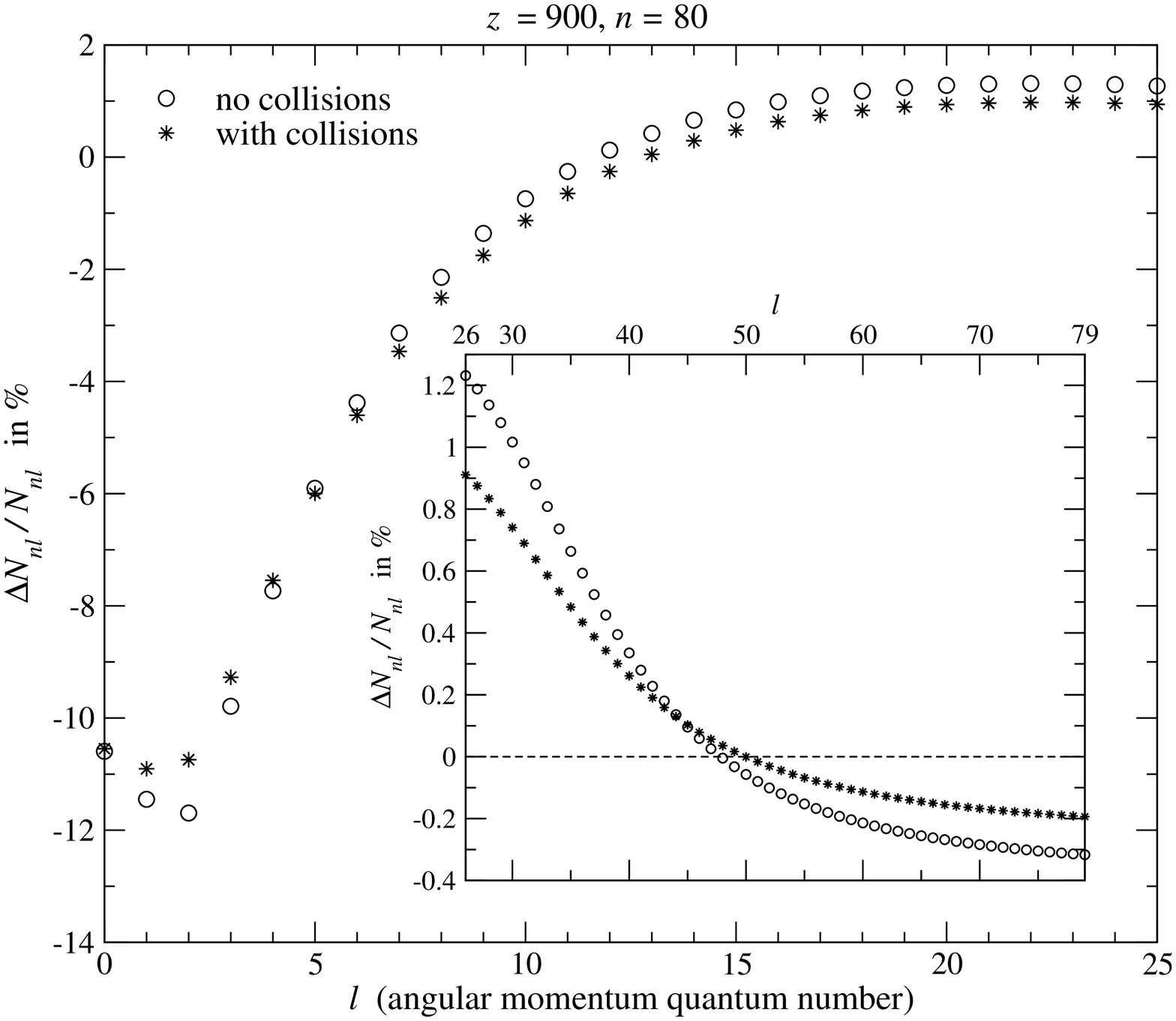}
\hspace{\hdist}
\includegraphics[width=\pscale\columnwidth]{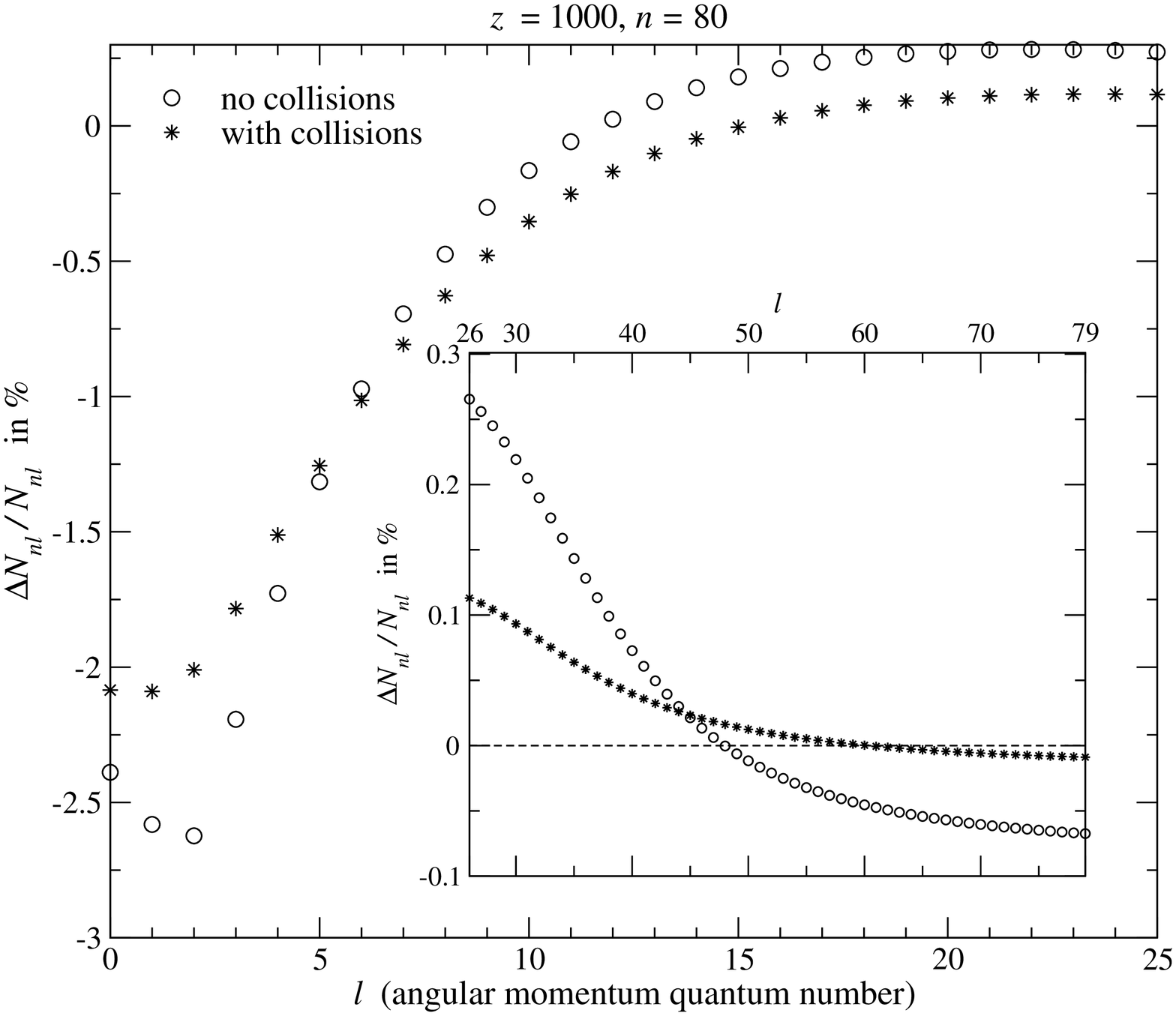}
\hspace{\hdist}
\includegraphics[width=\pscale\columnwidth]{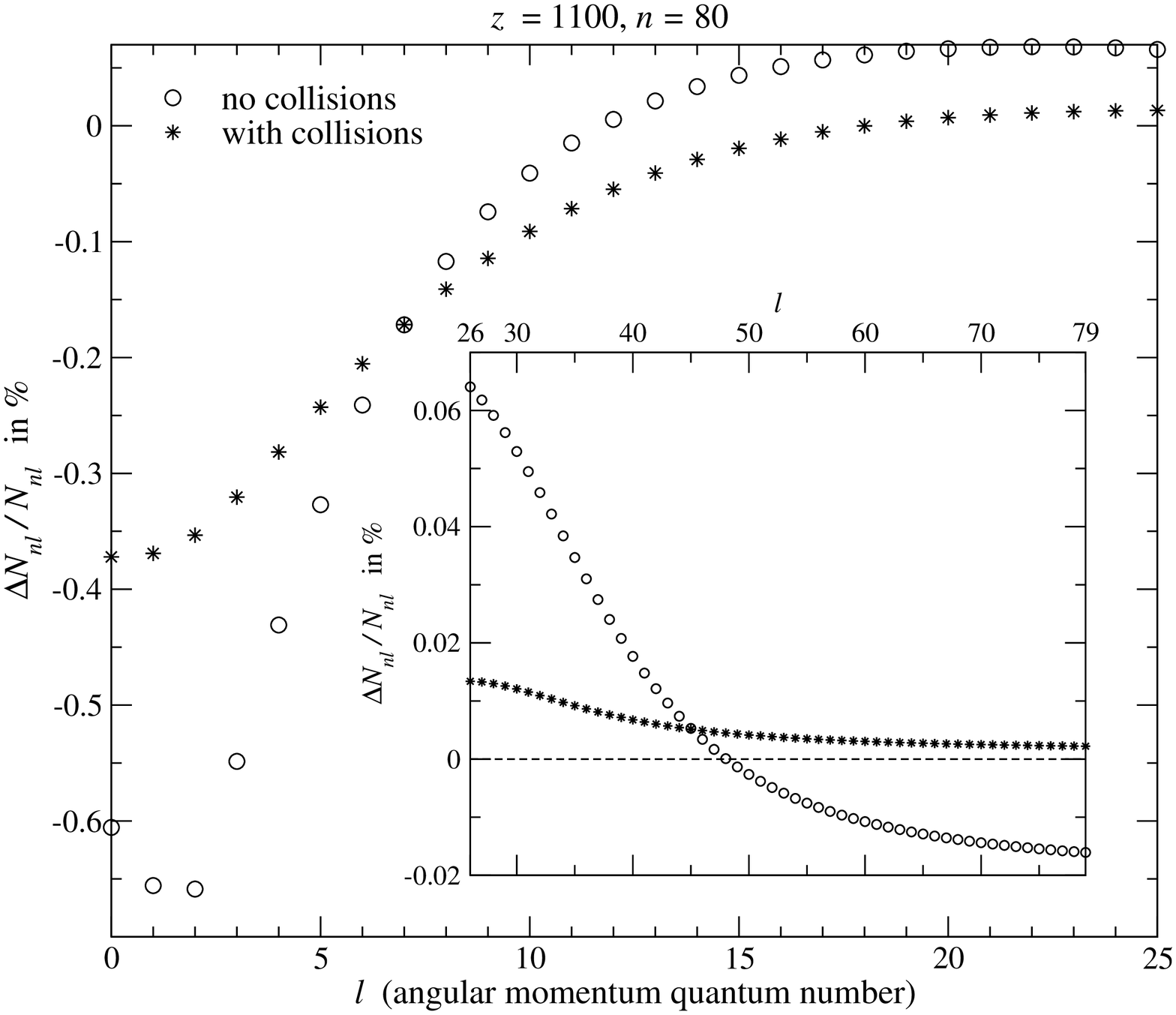}
\\[\vdist]
\includegraphics[width=\pscale\columnwidth]{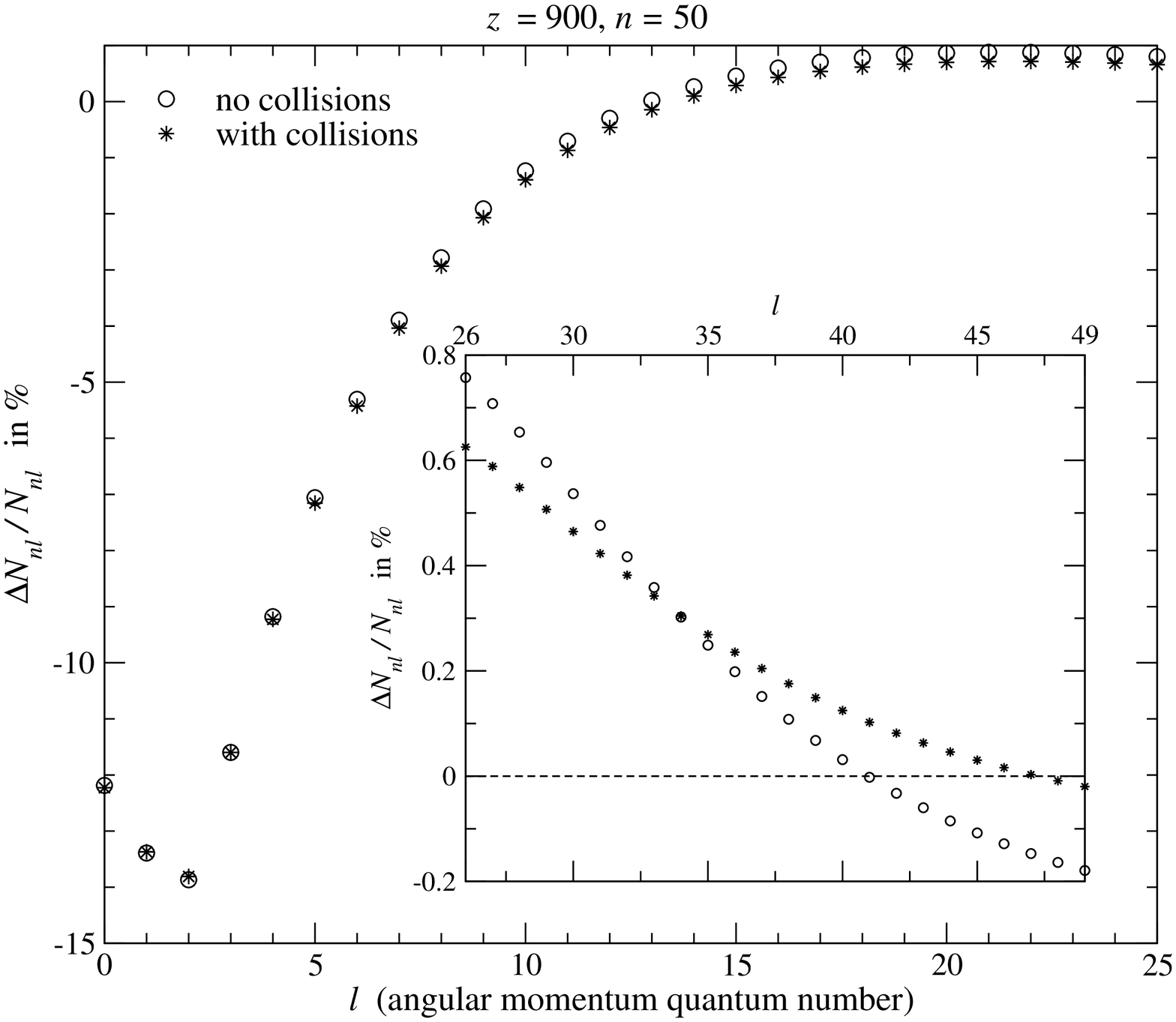}
\hspace{\hdist}
\includegraphics[width=\pscale\columnwidth]{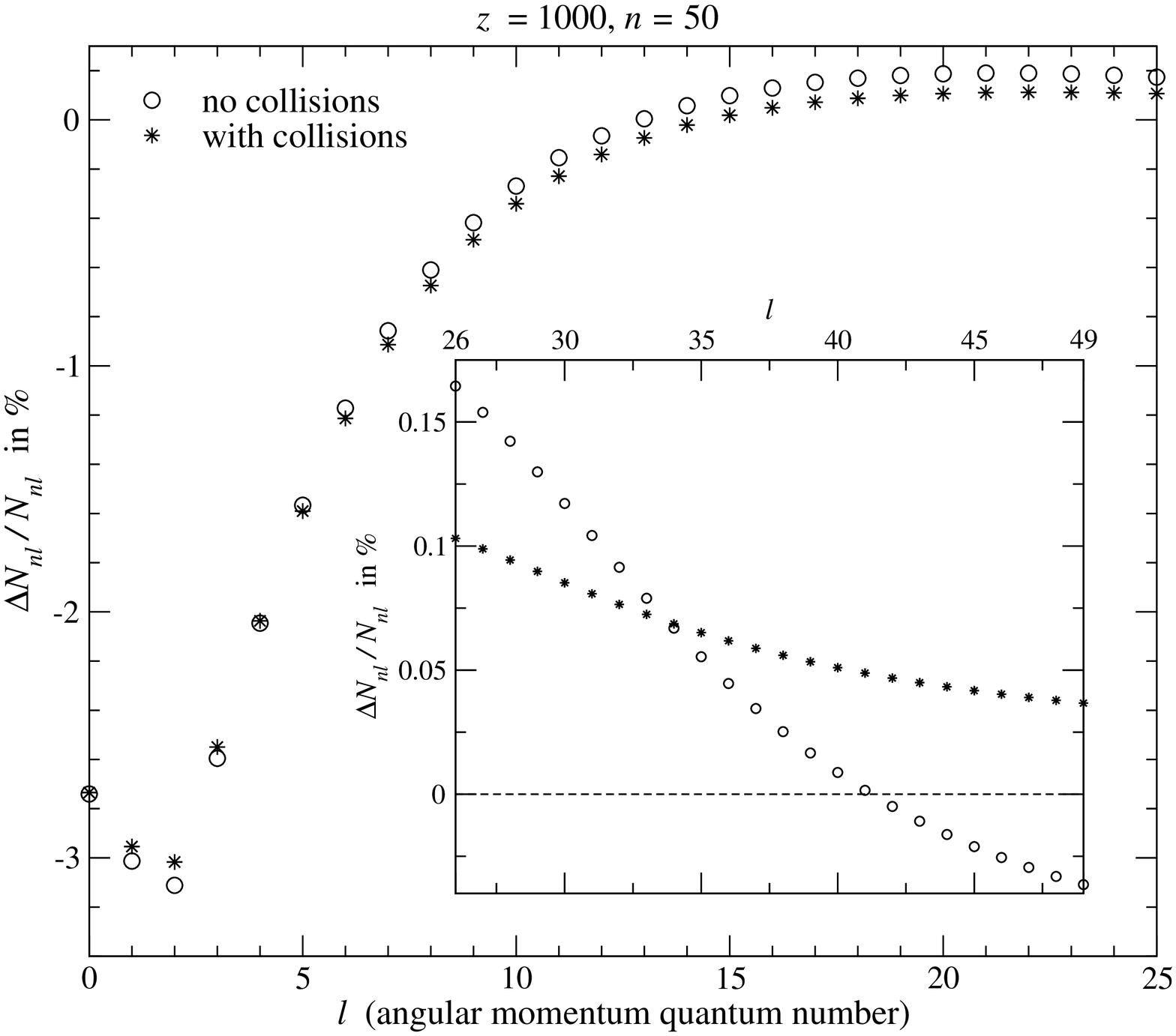}
\hspace{\hdist}
\includegraphics[width=\pscale\columnwidth]{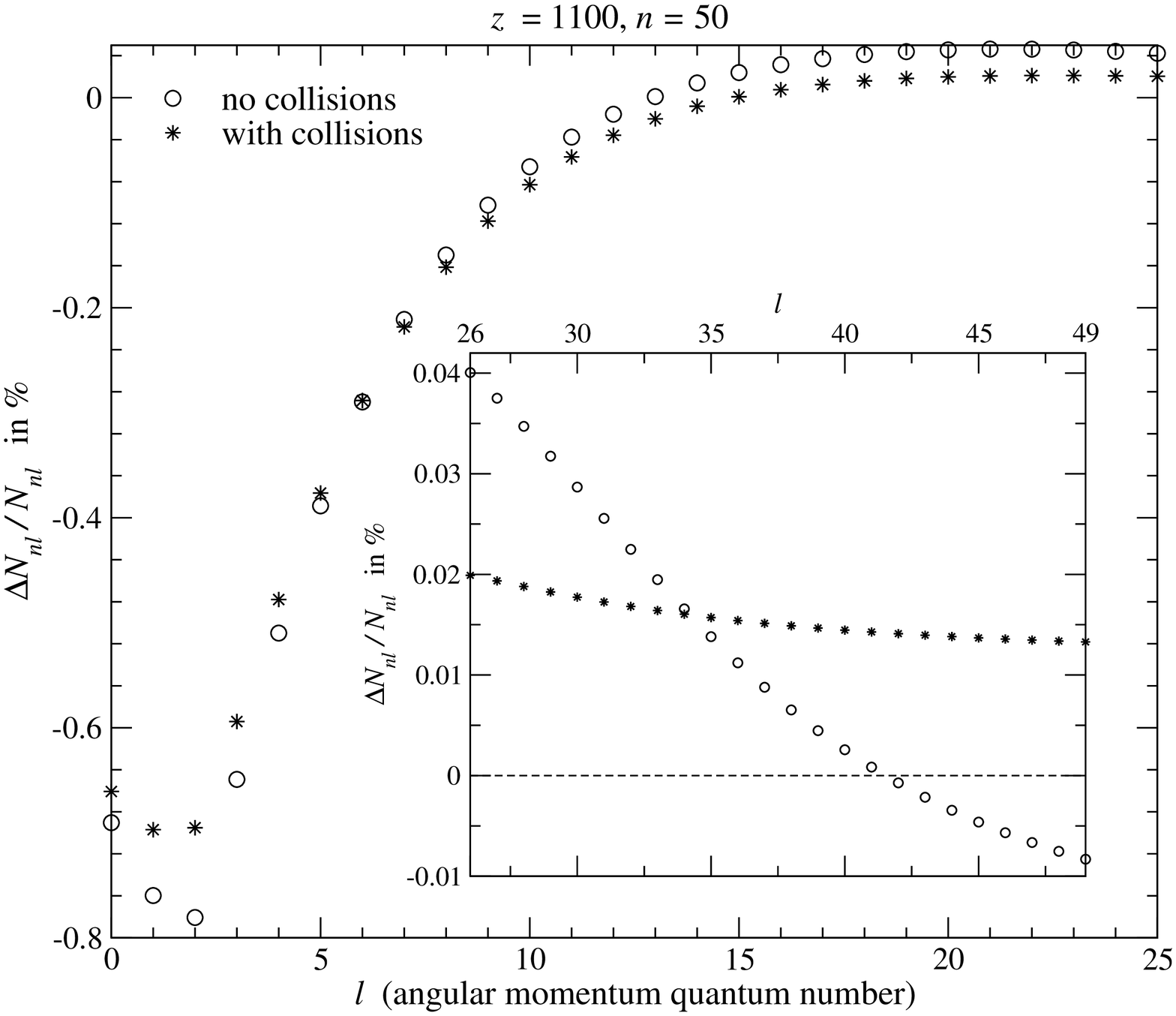}
\\[\vdist]
\includegraphics[width=\pscale\columnwidth]{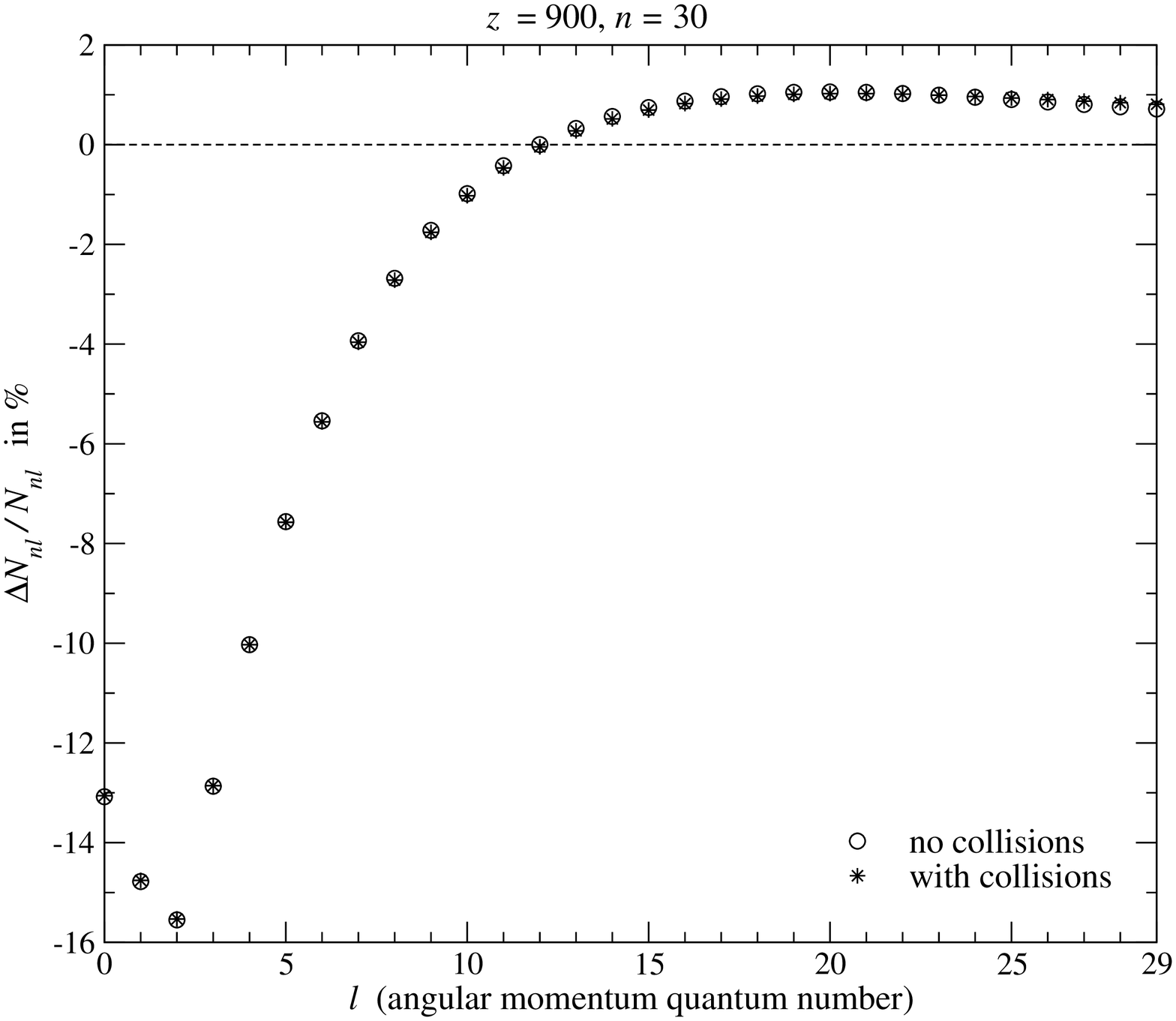}
\hspace{\hdist}
\includegraphics[width=\pscale\columnwidth]{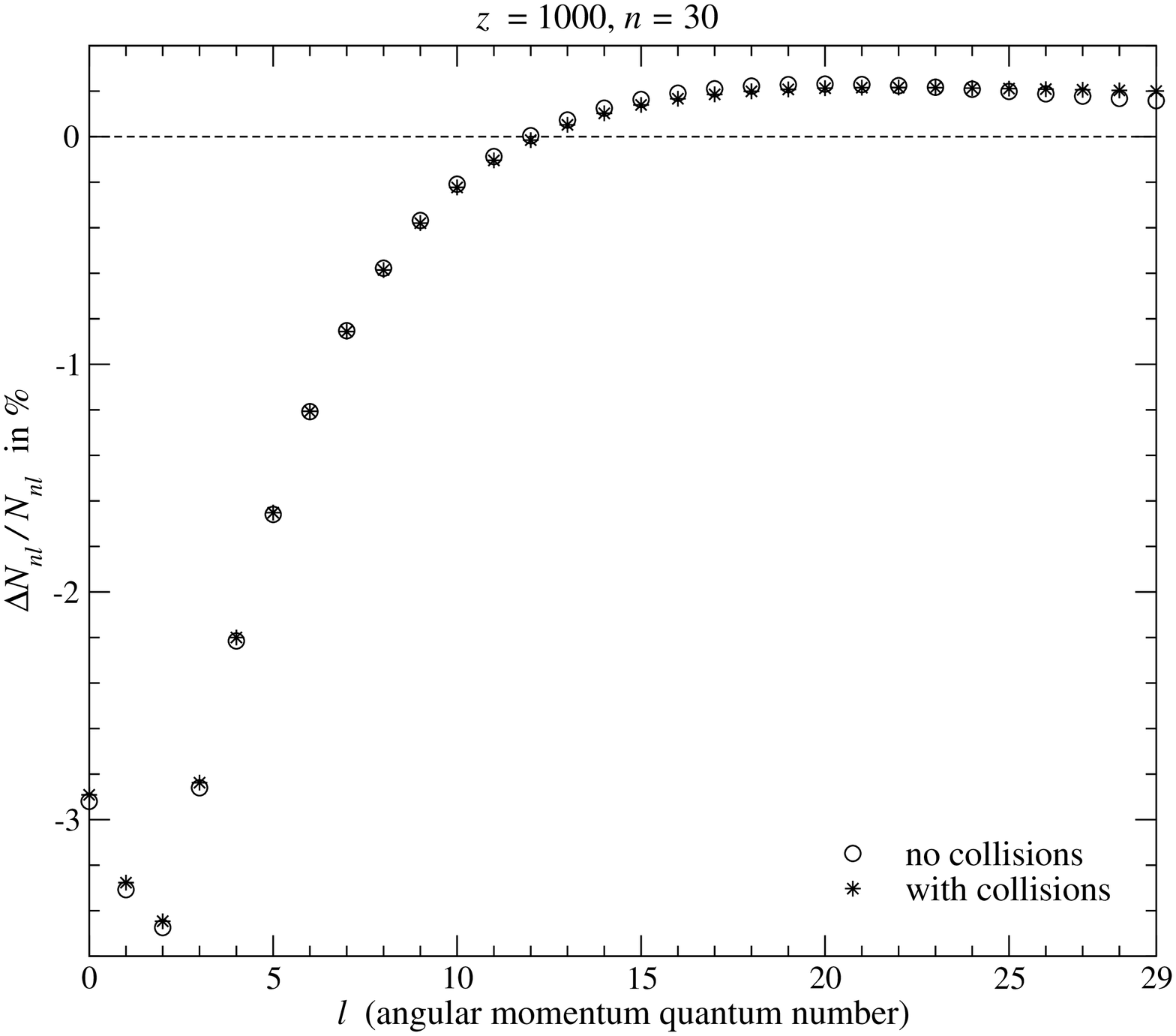}
\hspace{\hdist}
\includegraphics[width=\pscale\columnwidth]{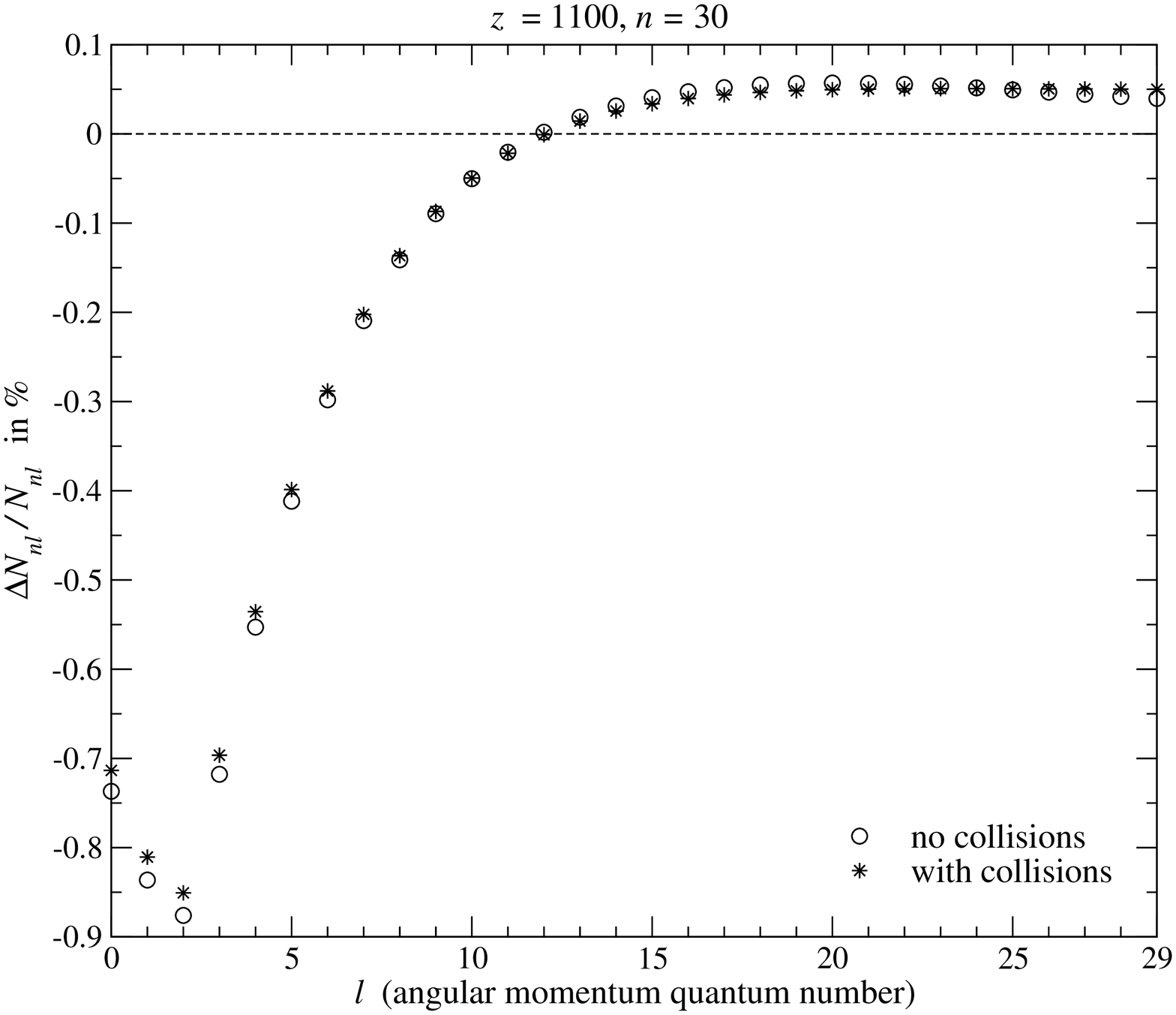}
\caption{Non-equilibrium effects on the populations of the angular momentum
sub-states for our computation with $\nmax=\nsplit=100$ at different redshifts
and given $n$. The left column is for $z=900$, the middle for $z=1000$ and the
right for $z=1100$. The computations were performed for $\nmax=100$ following
the populations of all the angular momentum sub-states separately. In each
panel the results obtained with and without the inclusion of $l$- and
$n$-changing collisions are given.
  We present $\Delta N_{n l}/N_{n l} \equiv [N_{n l}-N^{\rm SE}_{n l}]/N^{\rm
    SE}_{n l}$, where the statistical equilibrium (SE) population is computed
  from the actual total population of the shell by $N^{\rm SE}_{n l} =
  [(2l+1)/n^2] N_{\rm tot}$.  }
\label{fig:DI_SE_comp_low_z}
\end{figure*}

\begin{figure*}
\centering 
\includegraphics[width=\pscale\columnwidth]{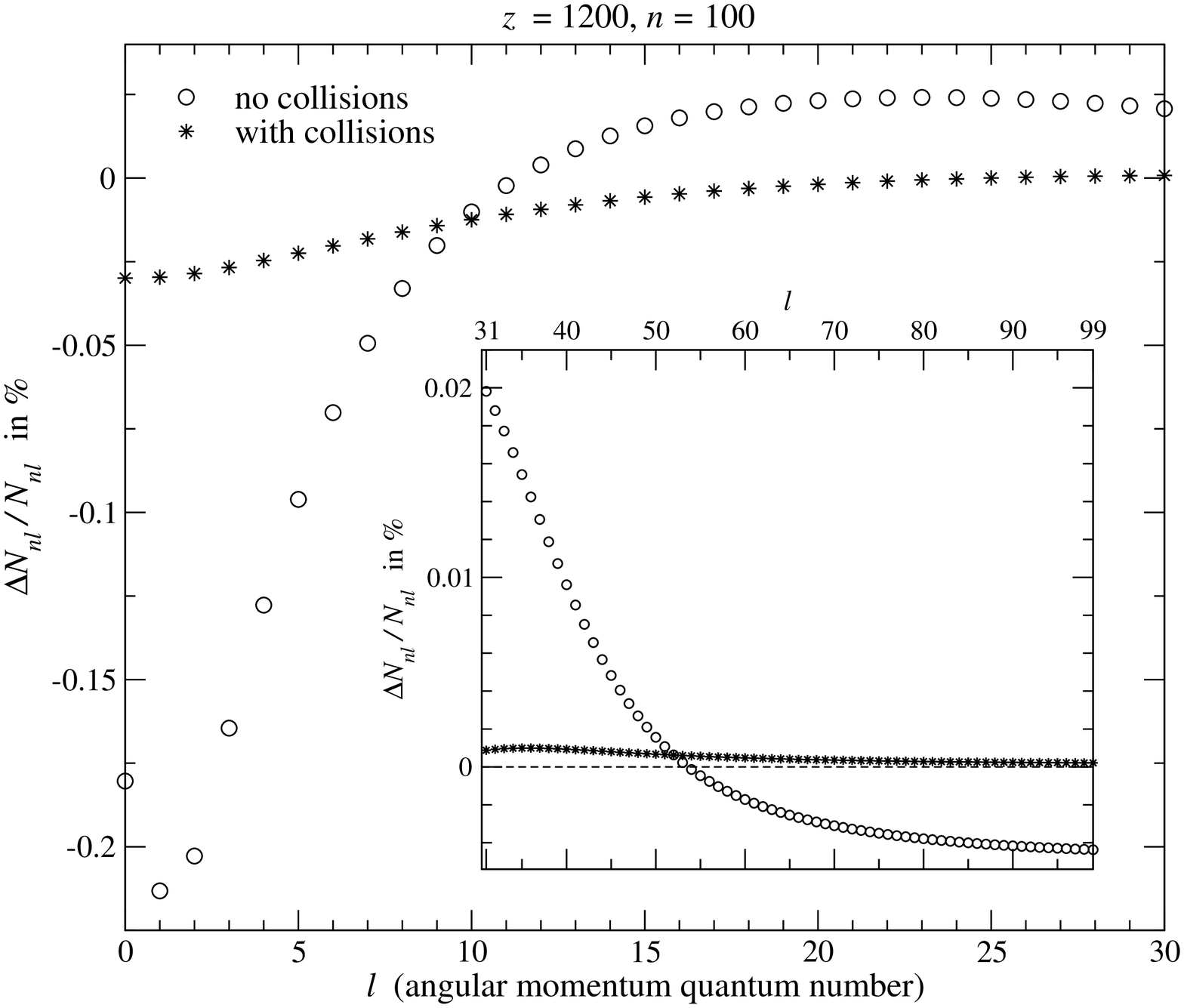}
\hspace{\hdist}
\includegraphics[width=\pscale\columnwidth]{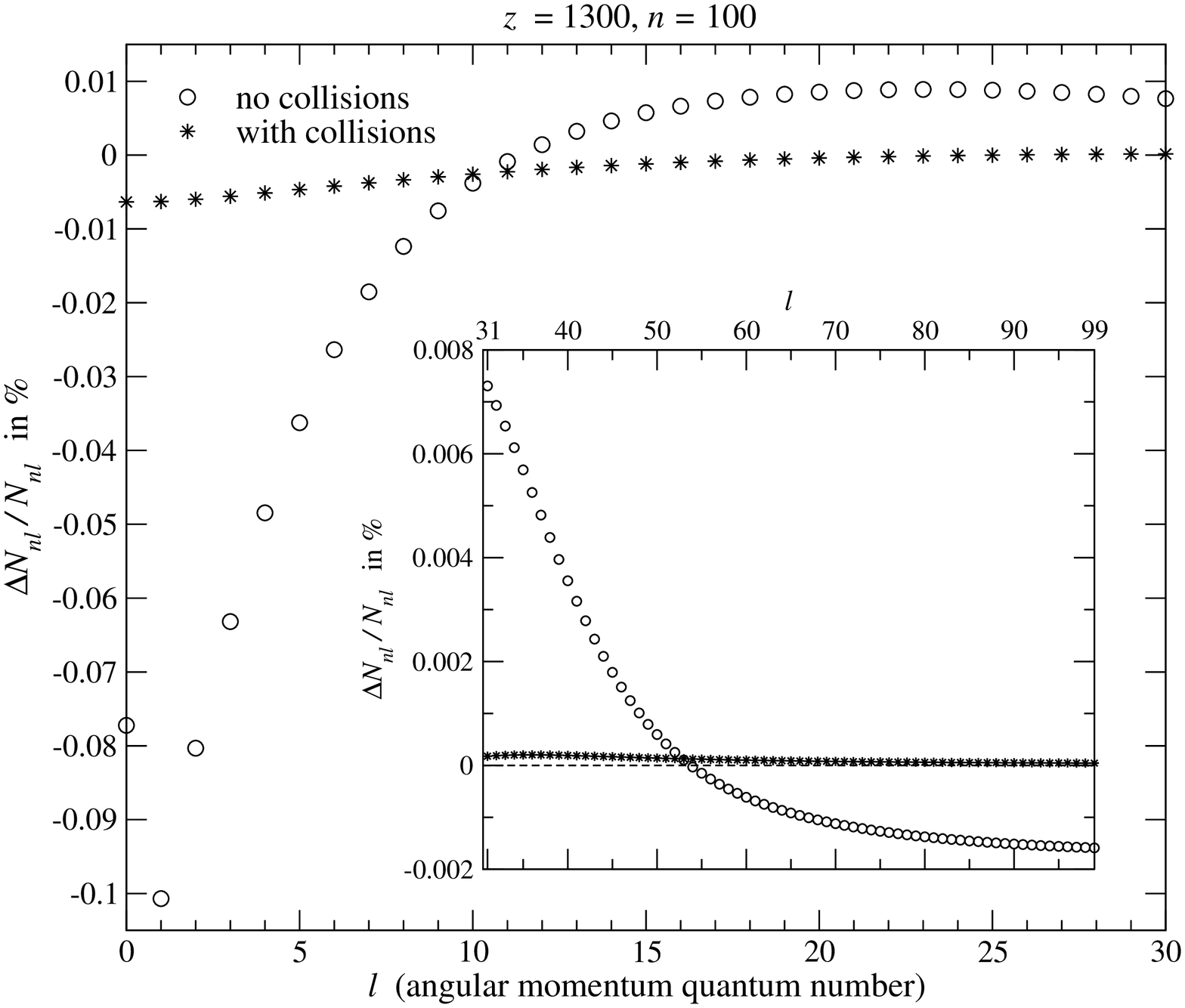}
\hspace{\hdist}
\includegraphics[width=\pscale\columnwidth]{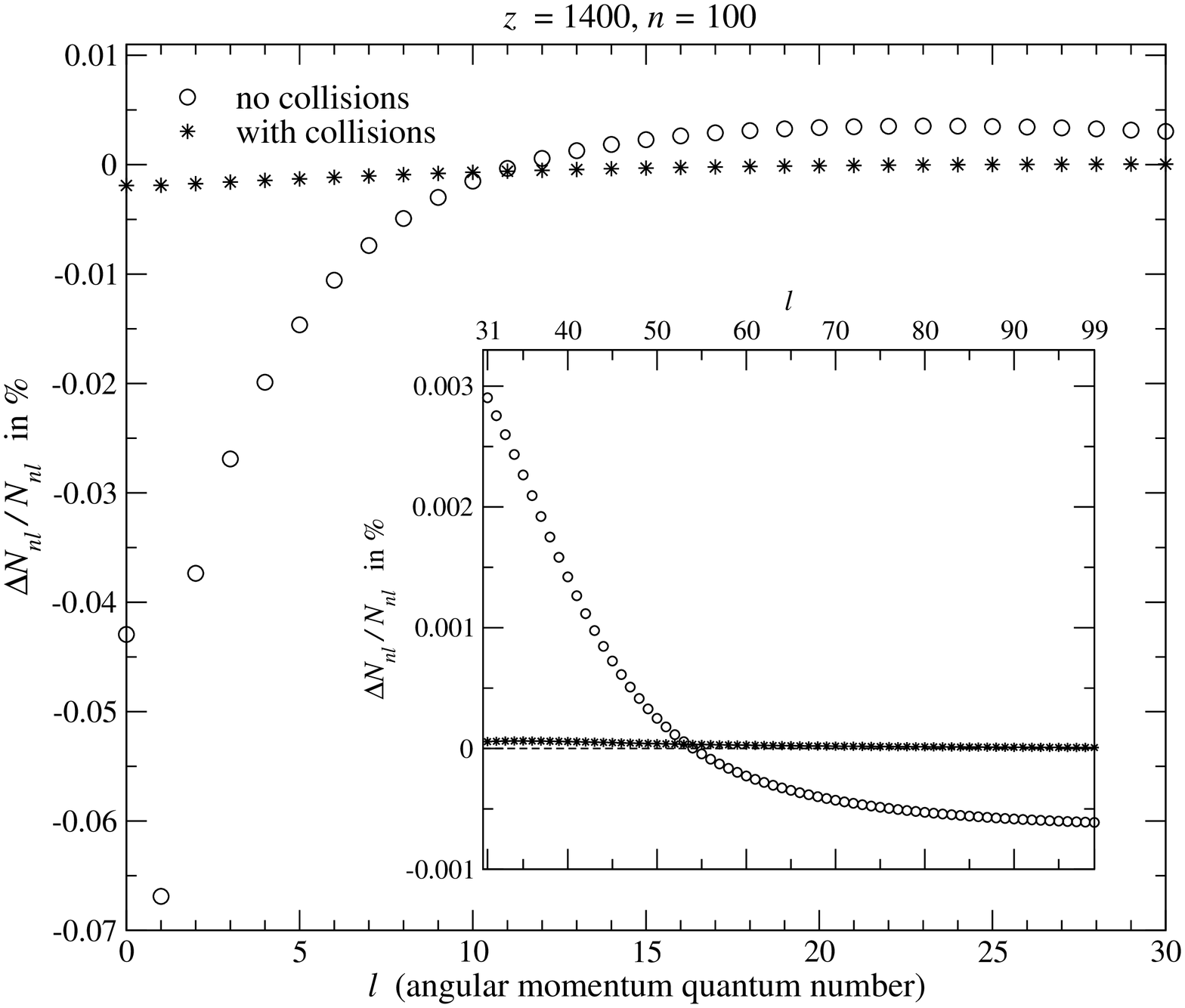}
\\[\vdist]
\includegraphics[width=\pscale\columnwidth]{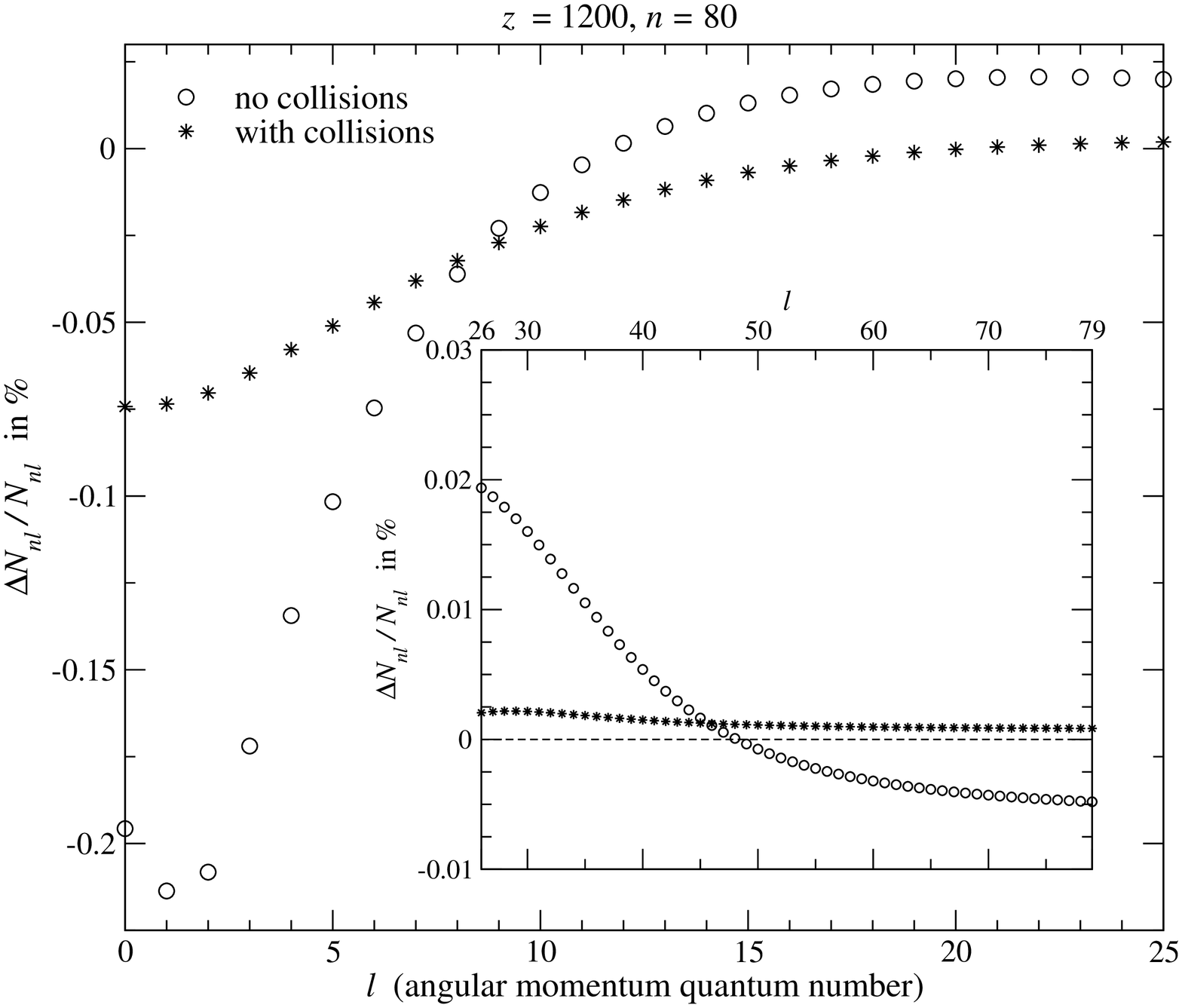}
\hspace{\hdist}
\includegraphics[width=\pscale\columnwidth]{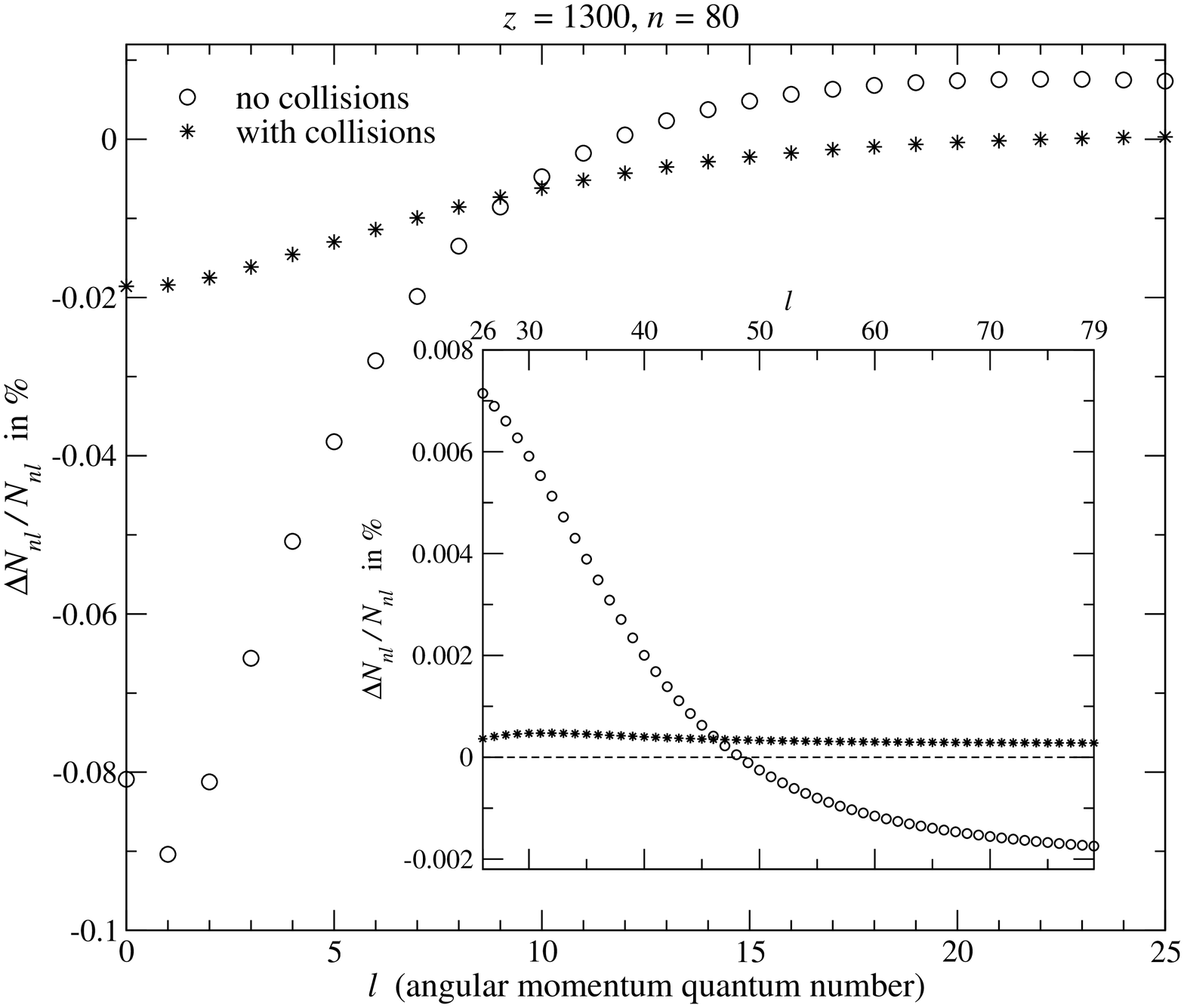}
\hspace{\hdist}
\includegraphics[width=\pscale\columnwidth]{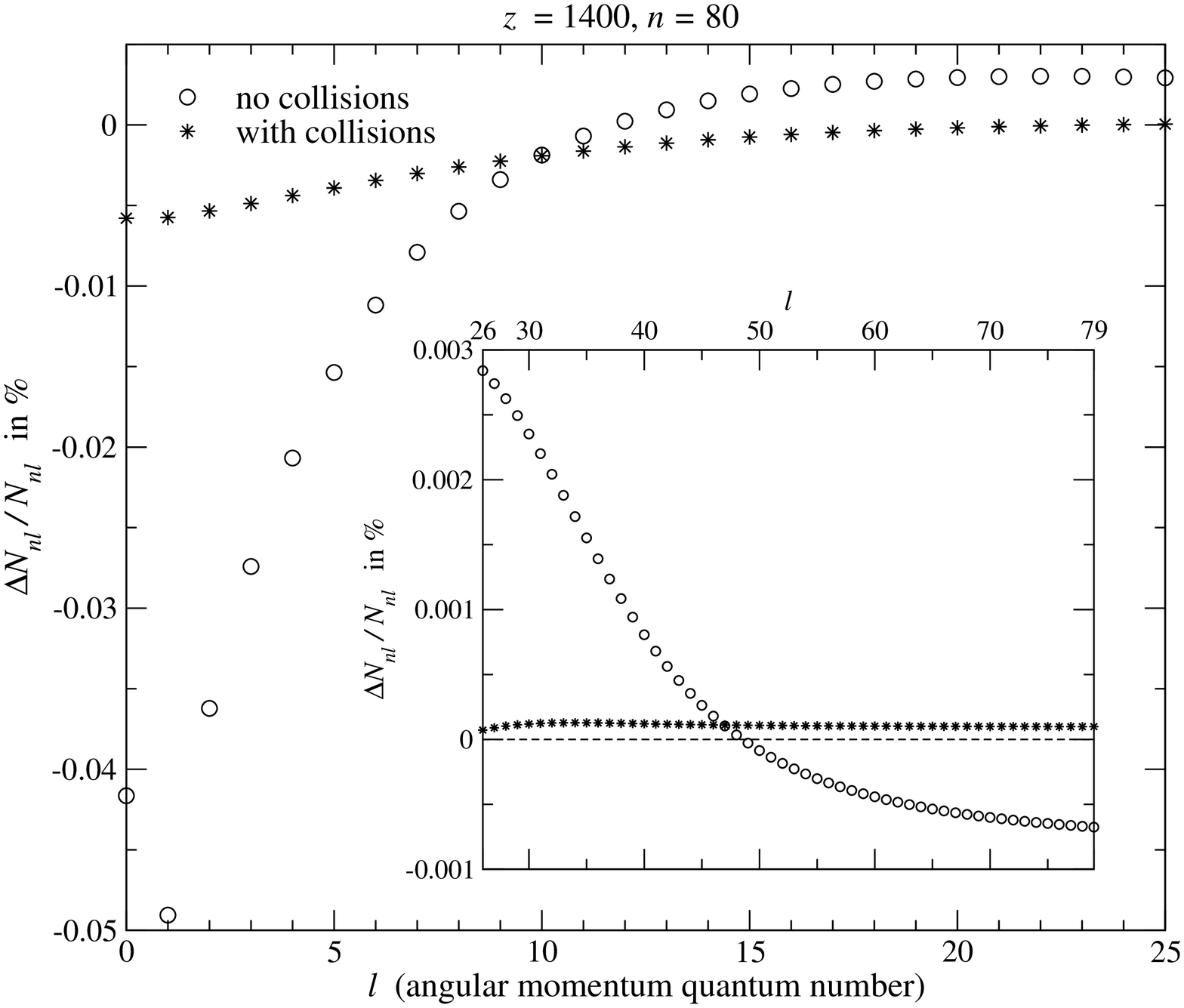}
\\[\vdist]
\includegraphics[width=\pscale\columnwidth]{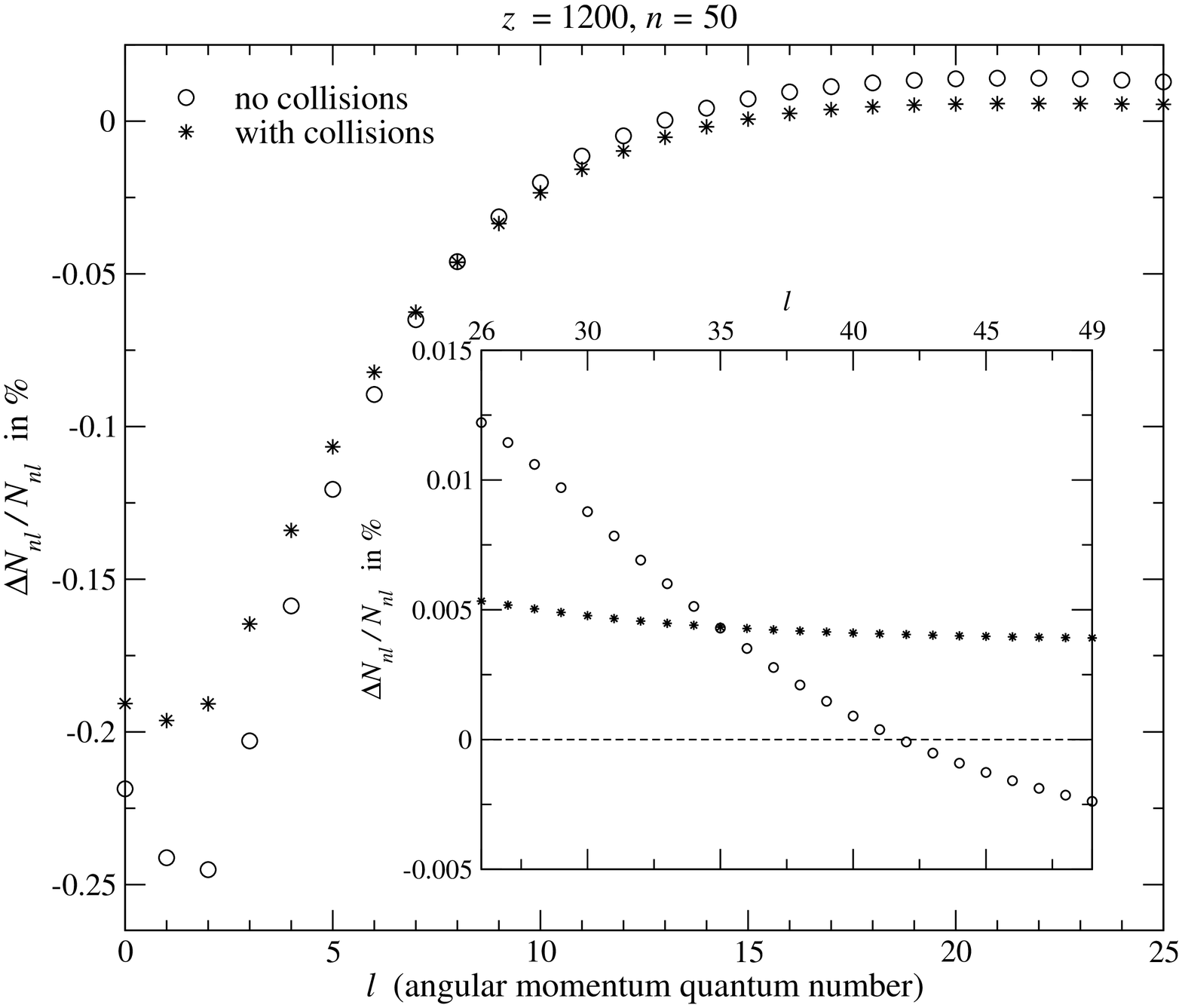}
\hspace{\hdist}
\includegraphics[width=\pscale\columnwidth]{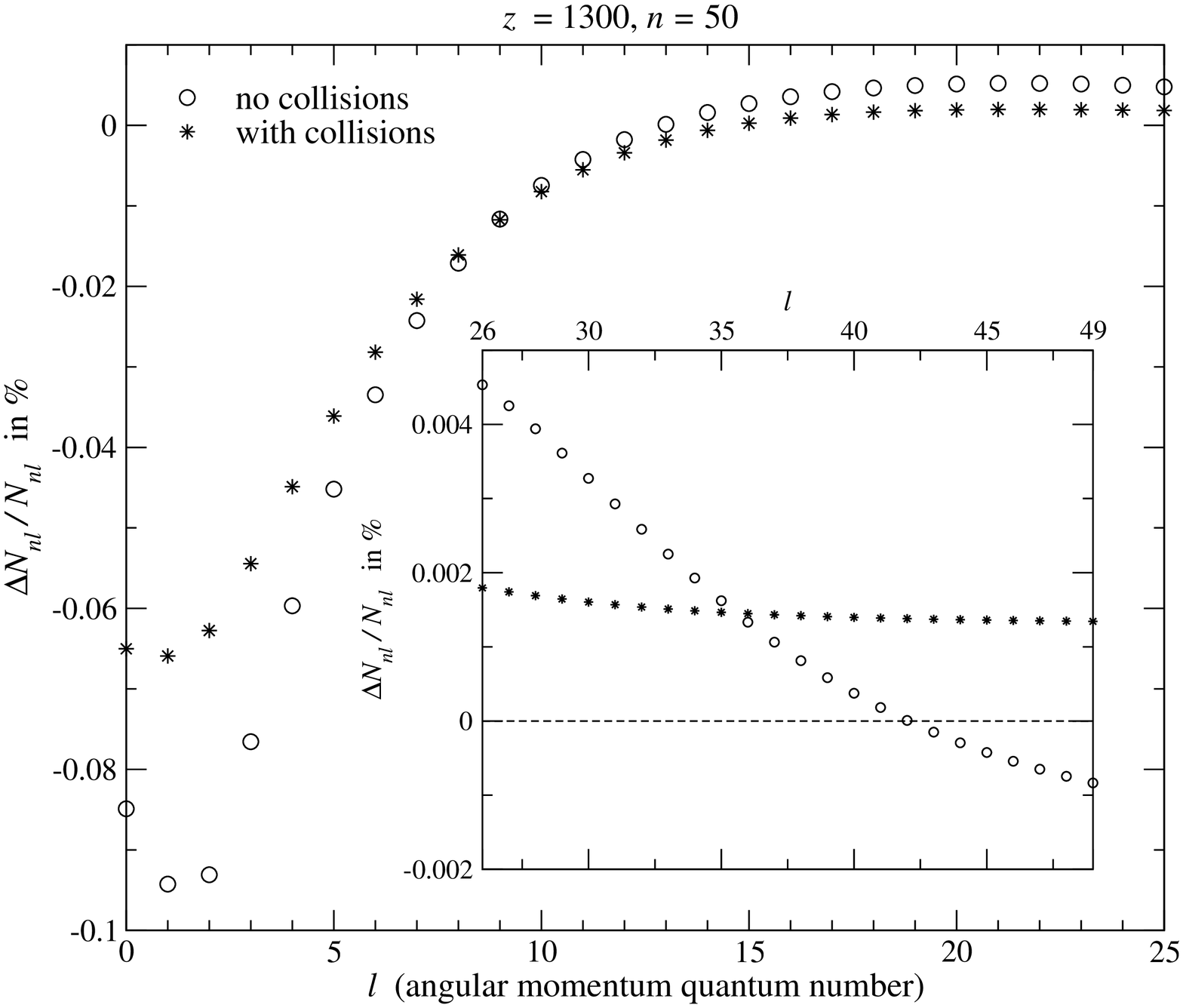}
\hspace{\hdist}
\includegraphics[width=\pscale\columnwidth]{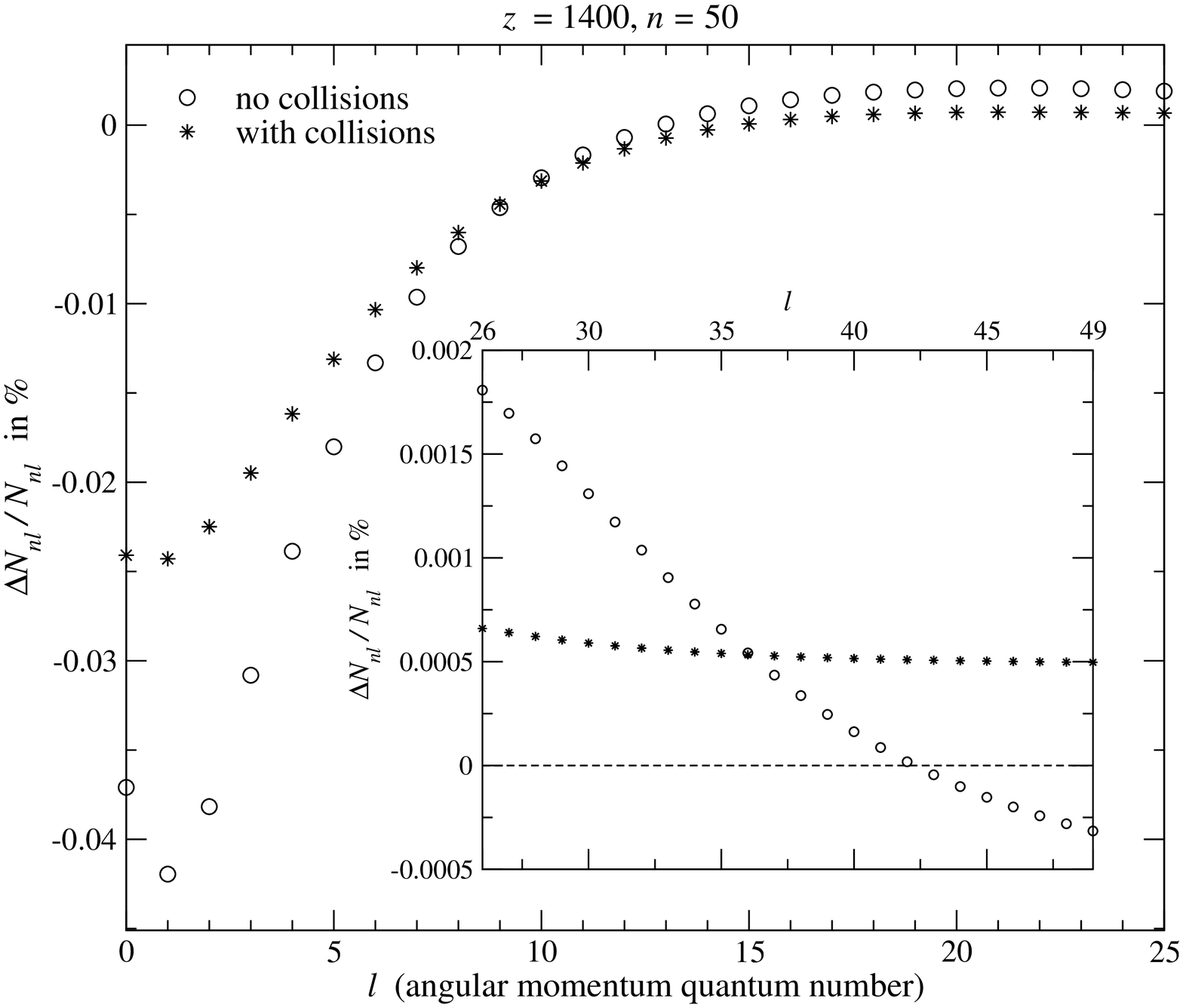}
\\[\vdist]
\includegraphics[width=\pscale\columnwidth]{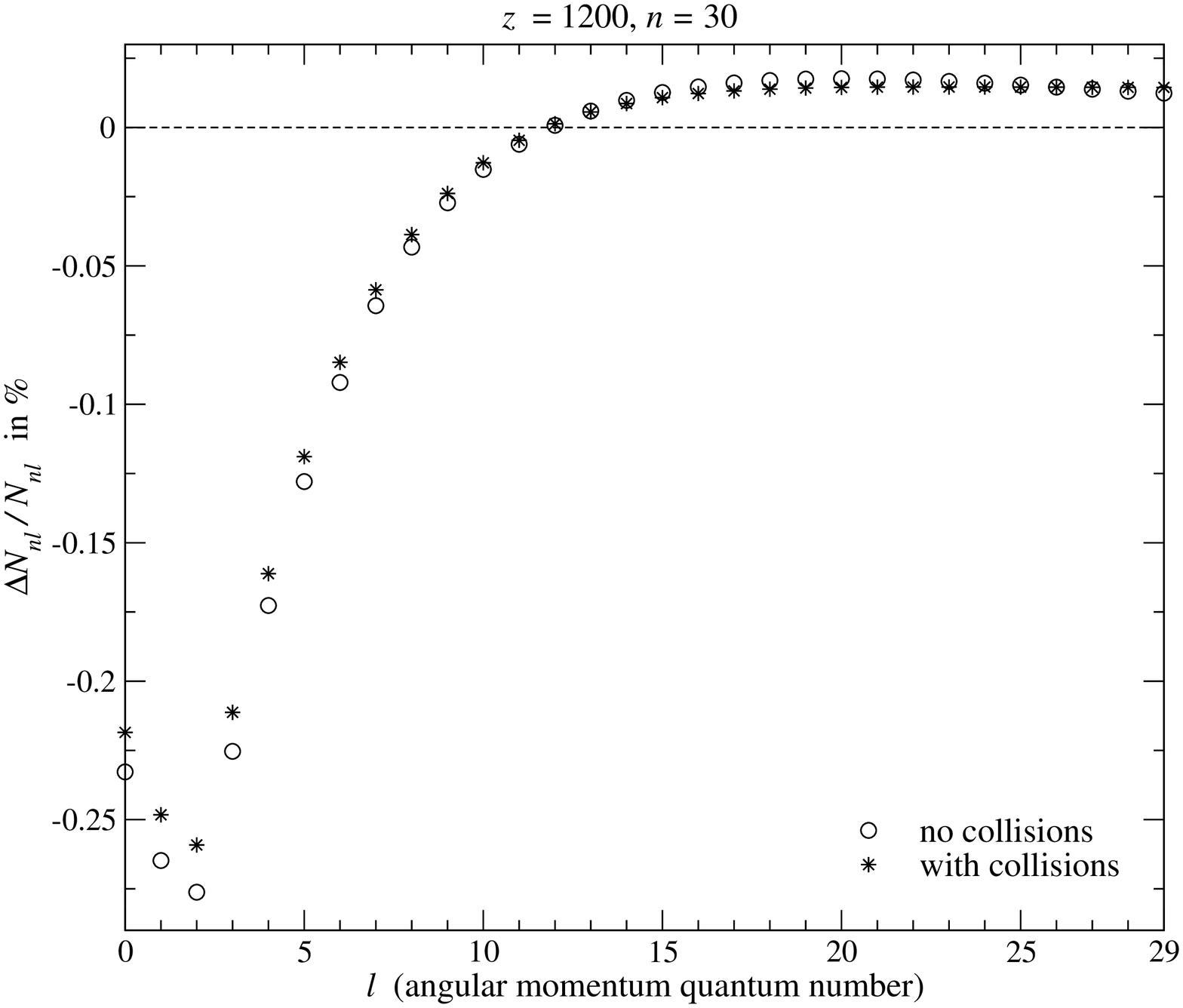}
\hspace{\hdist}
\includegraphics[width=\pscale\columnwidth]{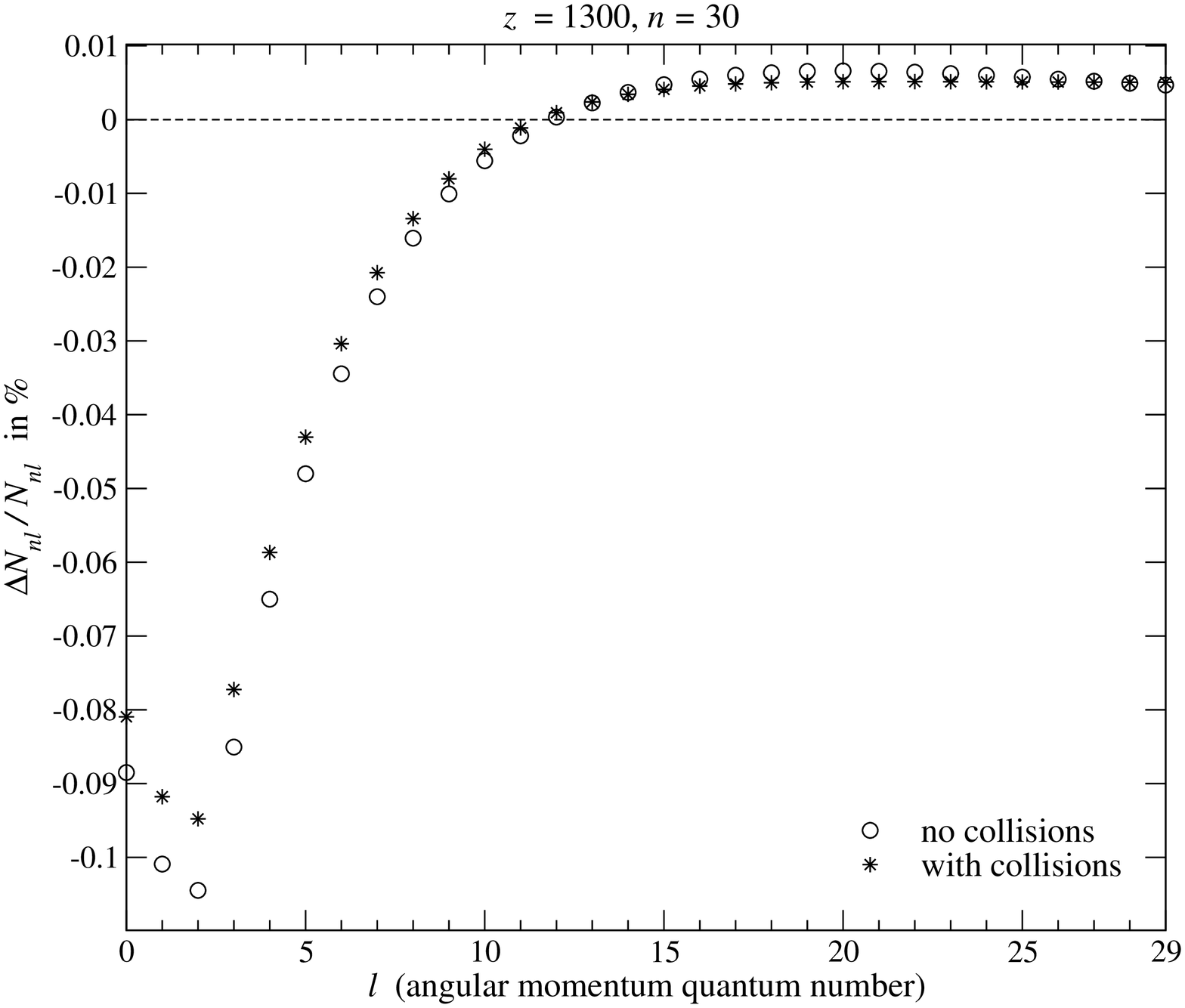}
\hspace{\hdist}
\includegraphics[width=\pscale\columnwidth]{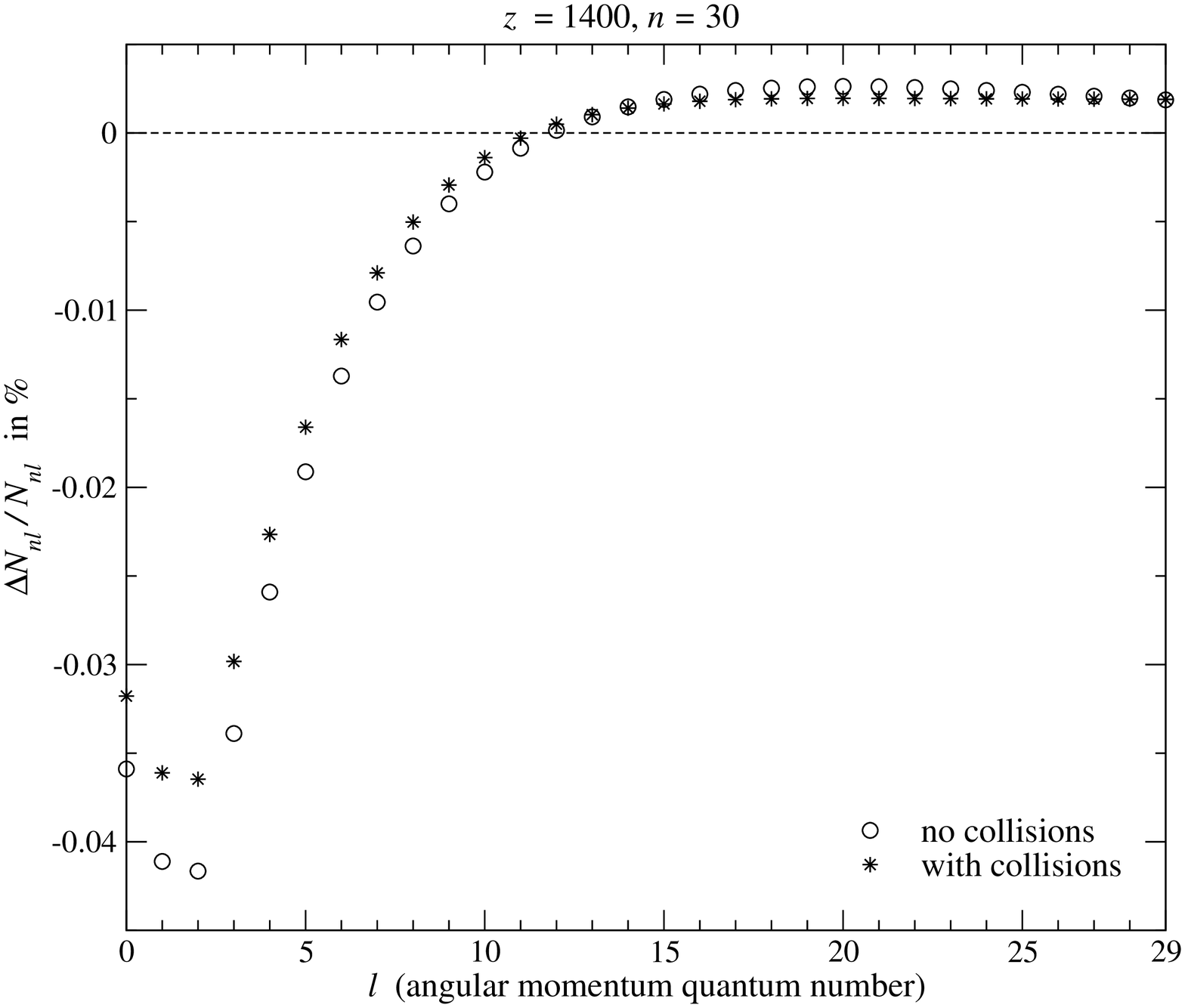}
\caption{ 
Non-equilibrium effects on the populations of the angular momentum sub-states
for our computation with $\nmax=\nsplit=100$ at different redshifts and given
$n$. The left column is for $z=1200$, the middle for $z=1300$ and the right
for $z=1400$. The computations were performed for $\nmax=100$ following the
populations of all the angular momentum sub-states separately. In each panel
the results obtained with and without the inclusion of $l$- and $n$-changing
collisions are given.
We present $\Delta N_{n l}/N_{n l} \equiv [N_{n l}-N^{\rm SE}_{n l}]/N^{\rm
SE}_{n l}$, where the statistical equilibrium (SE) population is computed from
the actual total population of the shell by $N^{\rm SE}_{n l} = [(2l+1)/n^2]
N_{\rm tot}$.  
%
}
\label{fig:DI_SE_comp}
\end{figure*}

In addition one can look at the deviations of the populations within a given
shell from full statistical equilibrium (SE) over the $l$ sub-levels.
In Figures \ref{fig:DI_SE_comp_low_z} and \ref{fig:DI_SE_comp} we
show the non-equilibrium effects on the populations of the angular momentum
sub-states at different redshifts and given $n$ for our computations
with $\nmax=\nsplit=100$. In each panel the results obtained {\it with} and
{\it without} the inclusion of $l$-changing and $n$-collisions are presented.
In general one can state that departures from full SE become smaller the
higher the redshift and the larger the value of $n$ is.
For $n=30$ the inclusion of collisions does not alter the results
significantly at {\it any} redshift, while for $n=50$ one can see that within
the high $l$-states the deviations from SE start to vanish.
For $n=80$ and $n=100$ within the high $l$-states the deviations from full SE
are very small down to redshifts $z\sim 1100-1200$ and even in the low
$l$-states departures from SE start to be wiped out.

%
But still even for $\nmax=100$ there is no shell beyond which full SE is
reached for {\it all} the angular momentum sub-states. All the low $l$-states
are underpopulated as compared to full SE. This is likely due to the fact that
for $n\gtrsim 40$ these levels can all connect more or less directly with
shells $n\lesssim 30$, which are always rather far from SE. Also the coupling
to the continuum due to the $l$-scaling of the photoionization Gaunt factor is
much stronger than for large $l$. Even for $n=100$ the $l$-changing collisions
are not able to significantly modify the low-$l$-populations set by the
radiative rates.
However, one still expects that increasing $\nmax$ for some $n>100$ also the
low $l$-states will reach full SE.  Unfortunately due to computational
constraints with $\nsplit=\nmax$ we were not able to go beyond $\nmax\sim
100$.
We conclude that for the cosmological hydrogen recombination calculations
$l$-changing collisions become important for shells with $n\gtrsim 30-40$, as
already estimated in RMCS06, but $n$-changing collisions may be neglected for
shells with $n\lesssim 100$.

\begin{figure}
\centering 
\includegraphics[width=0.96\columnwidth]{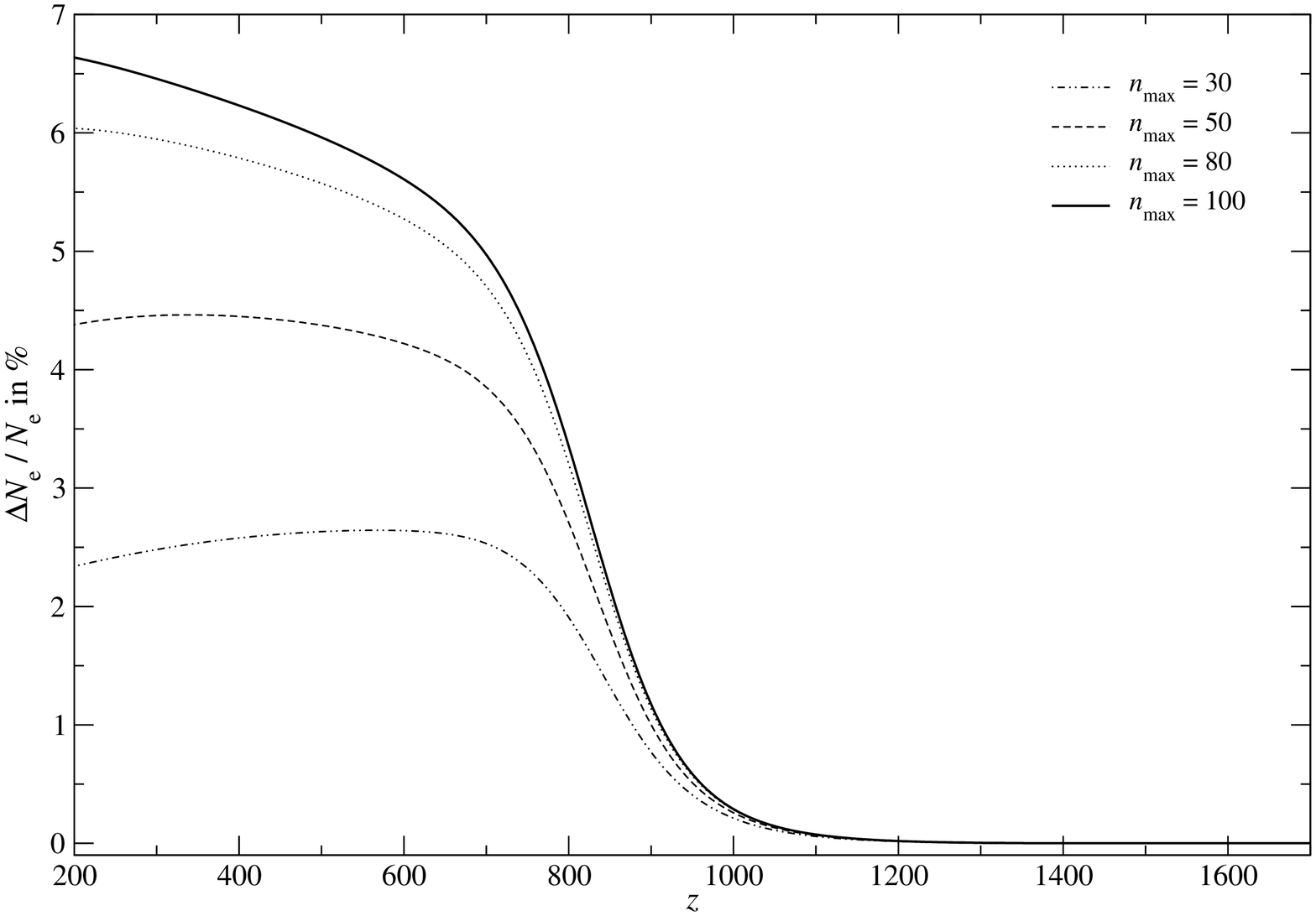}
\\
\includegraphics[width=0.96\columnwidth]{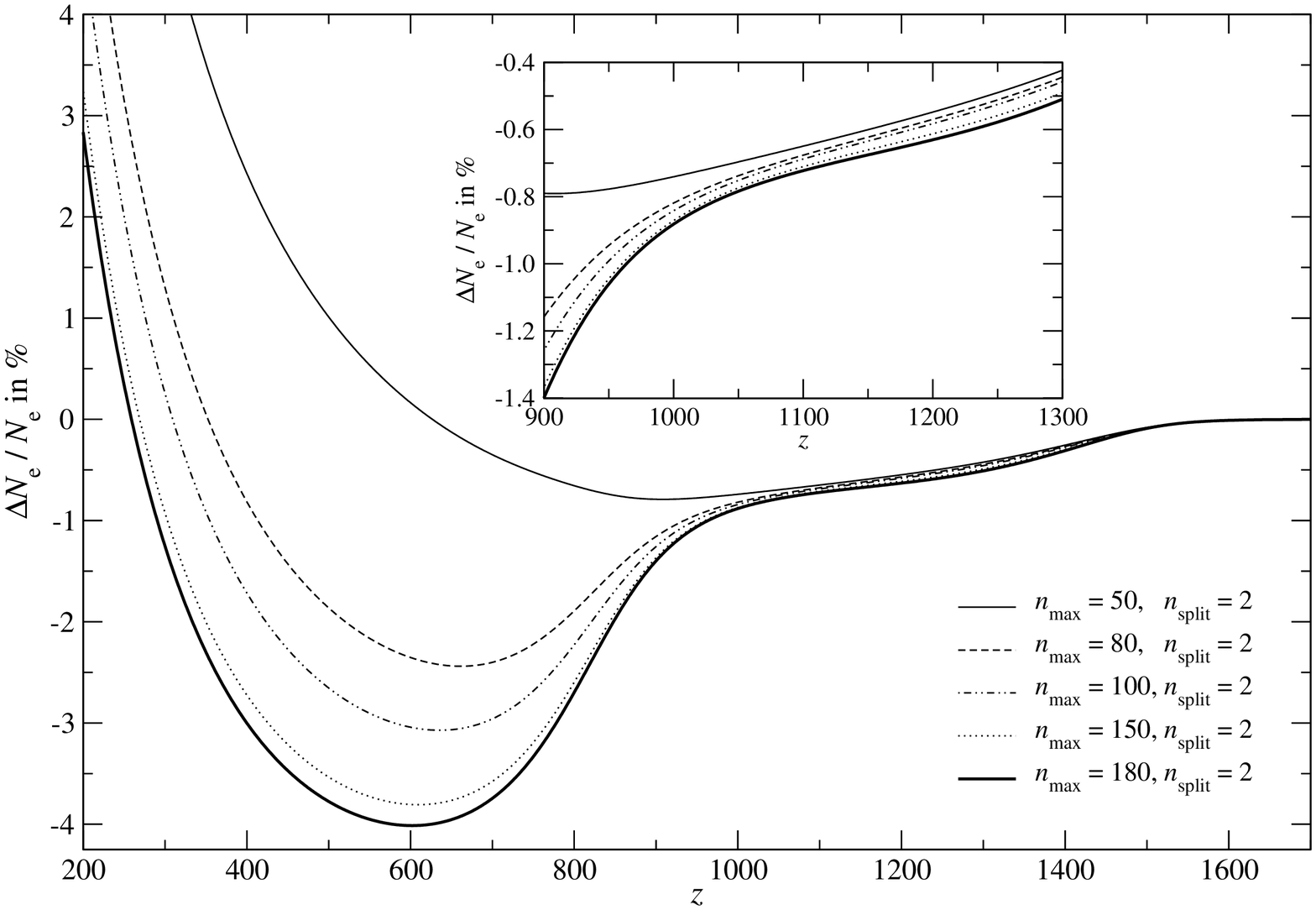}
\\
\includegraphics[width=0.96\columnwidth]{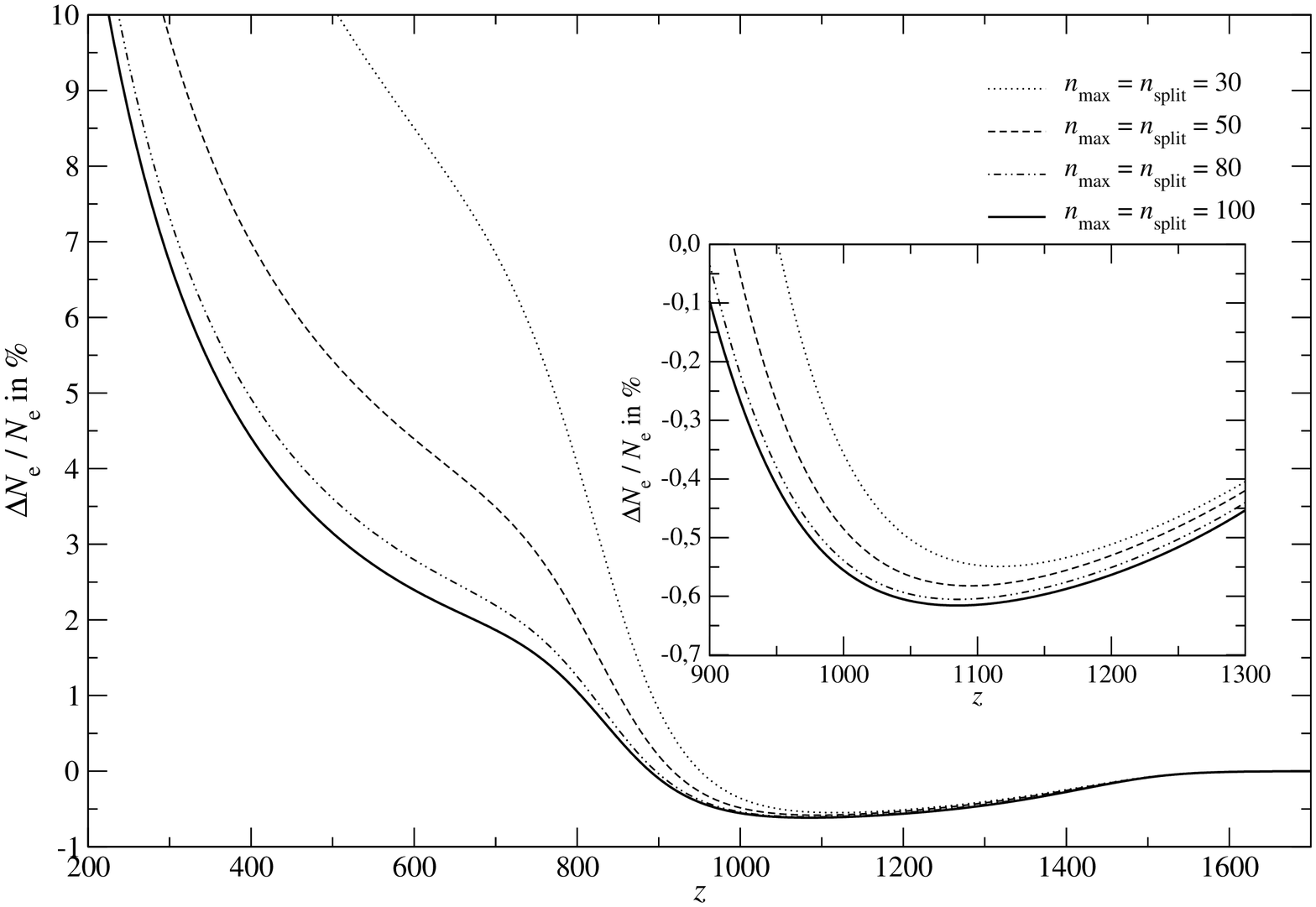}
\caption{Relative difference in the free electron fraction, $\Delta N_{\rm
e}/N_{\rm e}$, for different values of $\nmax$. The upper panel shows the
comparison for our computations with $\nsplit=\nmax$ with the results for
$\nsplit=2$. In the middle panel we compare our computations for $\nsplit=2$
with the output of {\sc Recfast} and in the lower panel we show the relative
difference for our computations with $\nsplit=\nmax$ with respect to the {\sc
Recfast} output.}
\label{fig:Xe_nmax_ns_2}
\label{fig:Xe_ns_2}
\label{fig:Xe_ns_max}
\end{figure}
\subsection{Effects on the free electron fraction and comparison with {\sc Recfast}}
\label{sec:el_hist}
In the upper panel of Fig.~\ref{fig:Xe_nmax_ns_2} we show the relative
difference in the free electron fraction for our computations with
$\nsplit=\nmax$ as compared to the results for $\nsplit=2$.
At low redshifts this difference exceeds the level of 5\%, \change{showing
that when following the populations of all the angular momentum sub-states
separately recombination is slightly slower. However, it seems that close to
the maximum of the Thomson visibility function ($z_{\rm dec}\sim 1090$ and
$\Delta z \sim 195$) departures from full SE within the shells are not
sufficiently large to modify the recombination history.}
At $z\sim 500$ the relative difference for the case $\nmax=100$ is more than 2
times \change{bigger} than for $\nmax=30$. Also in the redshift range
$z\gtrsim 700$ the ratio of these two curves remains greater than $\sim
1.2$. Including more than 100 shells into the calculation seems necessary in
order to reach full convergence at all redshifts.
In the middle panel of Fig.~\ref{fig:Xe_ns_2} we present the direct comparison
of our results with the output of {\sc Recfast}\footnote{\change{In fact we
    used our own implementation of this code, which solves the system of
    differential equations as given in \citep{SeagerRecfast1999} with accuracy
    better than 0.1\%.}}  for different cases with $\nsplit=2$ up to
$\nmax=180$. During the peak of the visibility function
%
%
the difference is close to $-0.7$\% and only reaches $\sim 1\%$ and $\sim 3\%$
for $z\sim 960$ and $z\sim 780$, respectively.
This comparison shows that representing the full multi-level calculation with
percent-level accuracy in the full range of redshifts probably requires some
more general 'fudge-function' instead of a single 'fudge-factor'.
\change{Also, the difference at $z\gtrsim 1000$ is probably not related to the
  detailed treatment of the populations in the angular momentum sub-states but
  points towards the limitations of the effective 3-level approach.}
A detailed analysis will be left for some future work.

Finally, in the lower panel of Fig.~\ref{fig:Xe_ns_max} we show the direct
comparison of our results for different calculations with $\nmax=\nsplit$ with
the output of {\sc Recfast}. There is a difference of roughly $-0.6$\% in the
electron fraction close to the peak of the visibility function.
The results also suggest that at low redshifts \change{recombination proceeds
  slightly slower than in the {\sc Recfast} code and that}
one can still expect more than a few percent corrections due to the detailed
treatment of the populations in the angular momentum sub-states.
However, one still has to push to a larger number of shells and very likely
include more physics, such as details of helium recombination, collisional
ionization and three-body recombination, collisional transitions with $|\Delta
l| > 1$ or forbidden transitions.
\change{At low redshifts ($z\lesssim 300$), also details of the hydrogen
  chemistry may have to be reconsidered.}
In addition one should include a simultaneous treatment of other processes
such as discussed by \citet{Dubrovich2005, Chluba2006, Kholu2006,
  Novosyadlyj2006} and likely more.
Here especially the treatment of two-photon transitions for higher shells
including stimulated emission or possibly other forbidden transitions may
still lead to percent level differences in both the hydrogen recombination
spectrum and the ionization history.

As an example, the strongest distortions of the CMB spectrum arising from the
epoch of hydrogen recombination are due to the Lyman-$\alpha$ transition and
the 2s two-photon decay.
Previously, the feedback of these excess photons onto the photoionization
rates \change{for} the second shell ($n=2$) has been considered by
\citet{Seager2000}, \change{yielding} no significant changes \change{in} the
ionization history.
However, recently it has been shown by \citet{Kholu2006} that this huge excess
of photons in the Wien-tail of the CMB leads to an increase of the ${\rm
  1s}\rightarrow{\rm 2s}$ two-photon absorption rate, which results in
percent-level corrections to the ionization history during hydrogen
recombination.
This feedback cancels a large part of the effect of induced two-photon decay
of the 2s level as discussed by \citet{Chluba2006} which again shows that
reaching percent-level accuracy for recombination is not very simple.

\change{We would like to point out that the simplification in connection with
  the computation of the photoionization and recombination rates, as described
  in Sect.  \ref{sec:addcomm}, at low redshifts may still affect the results
  presented in Fig.~\ref{fig:Xe_ns_2} on the percent-level.}

\begin{figure}
\centering 
\includegraphics[width=0.98\columnwidth]{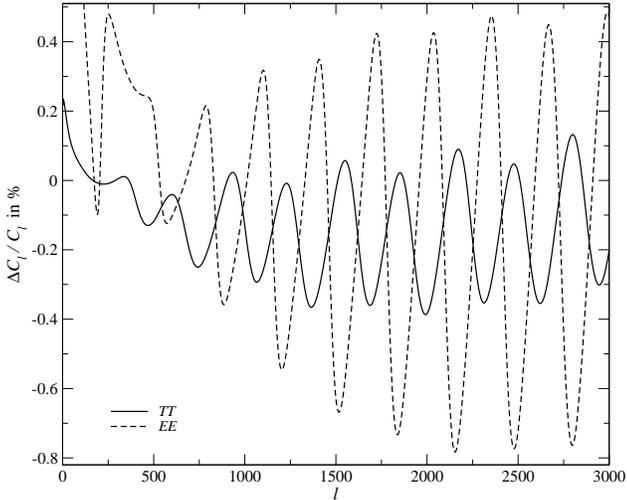}
\caption{changes of the CMB temperature ($TT$) and polarization ($EE$) power
spectra for our full computation with $\nmax=\nsplit=100$ including
collisions.}
\label{fig:DCl}
\end{figure}
\subsection{Changes in the CMB temperature and polarization power spectra}
\label{sec:DCl}
\change{In order to compute the changes of the CMB temperature and
polarization power spectra due to the differences in the ionization history we
modified the {\sc Recfast} routine inside the latest version of {\sc Cmbeasy}
\citep{Doran2005}. Since we obtained the ionization history only in the
redshift range $200\leq z \leq 2100$ we had to {\it merge} the solution of
{\sc Recfast} with our solution. At $z=2100$ this was not problematic since
there both solutions practically coincide. At $z\leq 200$ we simply rescaled
the {\sc Recfast} output by a constant factor, such that the merged solution
became continuous. We checked how this approach influences the obtained
difference in the power spectra by changing the transitions redshift to
$z=300$. However, this only affected the changes in the power spectra at
$l\lesssim 50-100$ on the level of percent and we expect that the results for
$l\gtrsim 100-200$ are robust.
\change{Note that here $l$ corresponds to the index of the spherical harmonic
function $Y_{lm}$ rather than the angular momentum quantum number for a level
inside the hydrogen atom.}

In Fig.~\ref{fig:DCl} we give the difference in the temperature and
polarization power spectra for our full computation with $\nmax=\nsplit=100$. The
difference reaches the level of 0.4\% for the $TT$ and 0.8\% for the $EE$ power
spectrum. We checked the convergence by using the results for different
$\nmax<100$ and found that the curves still changed at the percent-level when
going from $\nmax=80$ to $100$. This shows that still more shells should be
included and computations have to be extended down to $z=0$ in order to obtain
a definite answer.

}

\section{Additional physical processes}
\label{sec:addd}
\subsection{Effects of free-free absorption}
\label{sec:freefree}
The free-free process is increasingly important at low frequencies and should
eventually be able to wipe out any spectral distortion below some frequency
$\nu_{\rm ff}$.
Given the solutions for the recombination history and the emission resulting
from each bound-bound transition one can also take the free-free process into
account. \change{Since at redshifts $z\gtrsim 1200$ the difference in
  the electron and photon temperature is very small, for simplicity we assumed
  $\Te=\Tg=T_0(1+z)$ with todays CMB temperature $T_0=2.725\,$K.}
\begin{figure}
\centering \includegraphics[width=0.98\columnwidth]{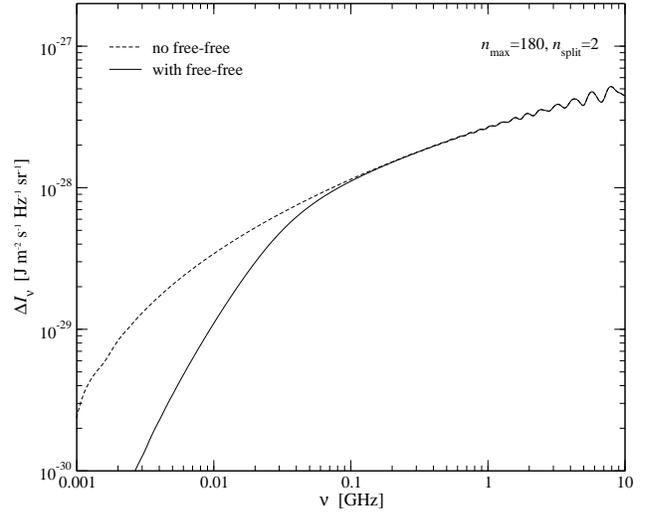}
\caption{Effect of free-free absorption on the hydrogen bound-bound emission
spectrum for $\nmax=180$ and $\nsplit=2$. The spectrum in the range
$\nu\gtrsim 100\,$MHz is not affected by the free-free process, whereas at
$\nu\lesssim 1\,$MHz all the distortions are erased.}
\label{fig:DI_ff}
\end{figure}
In Figure \ref{fig:DI_ff} we illustrate the effect of free-free absorption on
the hydrogen recombination spectrum for $\nmax=180$ and $\nsplit=2$. We chose
this particular example, since we could obtain the distortions down to
sufficiently low frequencies.
Our calculations show that the recombination spectrum in the frequency range
$\nu\gtrsim 100\,$MHz is not affected by the free-free process.
\change{However, for \ion{He}{iii} and probably \ion{He}{ii} recombination the
  effect of free-free absorption will be stronger, since the temperature and
  density of the plasma is larger and there are more electrons per helium atom
  when hydrogen is still fully ionized.}

\subsection{Effects of electron scattering}
\label{sec:el_sc}
To estimate the effects of electron scattering one can calculate the expected
broadening of an initially narrow line for a given recombinations history.
Following \citet{Pozdniakov1979} for large Thomson optical depth
$\tau=\int\Ne\sigT\id l\gg 1$ and small Compton y-parameter $y=\int
\frac{\kB\Te}{\me c^2} \id\tau\ll 1$ the broadening due to electron scattering
is dominated by the Doppler effect and one can find:
$\left.\frac{\Delta\nu}{\nu}\right|_{\rm sc}\sim 2 \sqrt{y\,\ln{2}}$.
Using the standard ionization history as obtained with the {\sc Recfast} code
one finds that for the spectral distortions due to hydrogen recombination the
broadening is less than percent for emission occurring at redshift $z\lesssim
1500$.  Since the feature we are considering here are typically much broader
($\gtrsim 20\%$) one may neglected the effect of electron scattering on the
hydrogen recombination spectrum.
However, for \ion{He}{ii} and in particular for \ion{He}{iii} recombination
this may not be true anymore due to two reason: first \ion{He}{iii} recombines
at higher redshift ($z\sim 6000-6500$) than hydrogen ($z\sim 1100-1500$) and
second because it occurred more rapidly. We shall not discuss these points any
further here.

\begin{figure}
\centering 
\includegraphics[width=0.98\columnwidth]{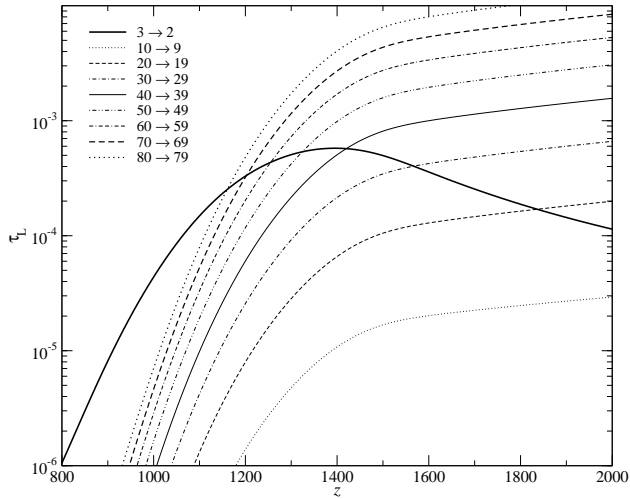}
\caption{Line optical depth for some $\alpha$-transitions within high
shells. The results were obtained using the computations for 100 shells. For
comparison the ${\rm H}_\alpha$-line ($3\rightarrow 2$) optical depth is given.}
\label{fig:Tau_S}
\end{figure}
\subsection{Imprint of the recombinational line transitions on the CMB angular power spectrum}
\label{sec:more}
Our detailed study of the hydrogen recombination permits us to evaluate the
effects discussed on \citet{Jose2005}, but now for shells above $n=10$.
In that paper, these authors discussed the imprint of the recombinational
lines on the angular power spectrum of the CMB, via resonant scattering in the
lines \citep[see also][for more details]{Kaustuv2004, Carlos2005}.
Due to correlations with the existing radiation field, the net effect on the
angular power spectrum, $\delta C_\ell$, is proportional to the optical depth
in the line, $\tau_{\rm L}$ \change{\citep[see Eq. 1 in][]{Jose2005}}, at the
relevant redshift.
As shown in \citet{Jose2005}, in the optically thin limit the first order
correction to the visibility function has two terms. The first one is
connected with {\it suppression} of existing anisotropies, and it is
proportional to $\tau_{\rm L} \times \mathcal{V}_{\rm T}$, where
$\mathcal{V}_{\rm T}$ is the Thomson visibility function \citep{Sunyaev1970}.
The second term is associated to {\it generation} of new anisotropies, and it
is proportional to $\tau_{\rm L} \times \exp{(\textstyle -\tau_{\rm T})}$.
\change{Note that here $l$ corresponds to the index of the spherical harmonic
function $Y_{lm}$ rather than the angular momentum quantum number for a level
inside the hydrogen atom.}

Since the bound-bound recombination lines are very narrow as compared to the
duration of recombination, fixing the observing frequency one tunes to the
signal arising during one particular epoch of recombination. Under the
assumption that the maximal signal for all the bound-bound recombination lines
arises at the same redshift this implies that going to lower frequencies the
effect is dominated by the signal coming from the $\alpha$-transitions of
higher shells.

In Fig.~\ref{fig:Tau_S} we present the optical depth to resonant scattering
for several main transitions ($\Delta n=1$) between high shells of the
hydrogen atom and we compare these to the one corresponding to ${\rm
  H}_\alpha$.
Although the optical depths for $n\gtrsim 20$ exceed the one for ${\rm H}_\alpha$ at
high redshifts, the expected signature in the power spectrum will be small due
to the presence of Thomson scattering.
However, this process might become interesting in the future because at low
frequencies the signals of many transitions overlaps and one expects that for a
given bandwidth of the experiment a superposition of the contributions with
comparable amplitude might amplify the signal. This may increase the total
change in the temperature power spectrum notably, especially at frequencies
below $\sim 1\,$GHz.
We have computed the effect on the power spectrum and found that for $\nu\sim
50-100\,$MHz it is comparable to the signal produced by the scattering within
the ${\rm H}_\alpha$-transition. Our computations also showed that a similar
process due to the {\it absorption} and {\it re-emission} of photons in the
continuum
%
%
\change{also leads to weak fluctuations on the level of $\delta C_\ell \ell (\ell
+1)/2\pi \approx 9\times 10^{-4} \mu {\rm K}^2$ at 1.4~GHz. This signal also shows a
characteristic oscillatory behaviour, and again it is maximum at small angular
scales ($\ell \sim 900$).}

\section{Conclusion}
We presented results for the spectral distortions of the CMB arising during
the epoch of cosmological hydrogen recombination due to bound-bound transition
with observing frequencies down to $\nu\sim 100\,$MHz and confirm that the
relative spectral distortions, $\Delta I_\nu/B_\nu$, are rapidly growing
towards low frequencies.
We show that even within the highest considered shell full statistical
equilibrium is not reached and that at low frequencies the recombination
spectrum is significantly different when full SE for shells with $n>2$ is
assumed.
We also conclude that in cosmological hydrogen recombination calculations
$l$-changing collisions should be included for shells with $n\gtrsim 30-40$,
but $n$-changing collisions can be neglected for shells with $n\lesssim 100$.

In Sect.~\ref{sec:el_hist} we have presented our results for the ionization
fraction as compared to the {\sc Recfast} output.
\change{Including a treatment of the populations of all the individual angular
  momentum sub-states slows recombination down, leaving a $\gtrsim 5-10\%$
  larger residual electron fraction at low redshifts. Although the overall
  effect on the temperature and polarization power spectra due to the changes
  in the recombination history is $\lesssim 1\%$ these additional electrons
  may be important in computations of the first stars and structures
  \citep{Barkana2001, Bromm2002}, for which especially the formation of
  molecular hydrogen plays a crucial role.}
The discussion has shown that representing the full multi-level calculation
with percent level accuracy in the full range of redshifts probably requires
some more general 'fudge-function' instead of a single 'fudge-factor'.
Also, one still has to push to a larger number of shells and very likely
include more physics.
For extensive calculations of the CMB temperature and polarization power
spectra it will become important to develop a new simplified and fast scheme
which accurately models the physics during the recombination epoch.

\section*{Acknowledgments}
We thank Moncef Derouich for useful discussions about collisional rates.  
We also acknowledge use of {\sc Cmbeasy} \citep{Doran2005}.
In addition J.C. and R.S would like to thank the Institute for Theory and
Computation of the Harvard University and the Institute of Advanced Study,
Princeton, for their hospitality during a visit in Sept.~2006 and useful
conversations at the final stages of this paper.

\bibliographystyle{mn2e}
\bibliography{Lit}

\label{lastpage}

\end{document}